# Title: The Least Difference in Means: A Statistic for Effect Size Strength and Practical Significance

**Authors:** Bruce A. Corliss[1,*], Yaotian Wang[2], Heman Shakeri[1], Philip E. Bourne[1,3]

**Affiliations:**

[1]School of Data Science, University of Virginia; Charlottesville, Virginia

[2]Department of Statistics, University of Pittsburgh; Pittsburgh, Pennsylvania

[3]Department of Biomedical Engineering, University of Virginia; Charlottesville, Virginia

*Corresponding author. Email: bac7wj@virginia.edu

**Abstract:**

With limited resources, scientific inquiries must be prioritized for further study, funding, and translation based on their practical significance: whether the effect size is large enough to be meaningful in the real world. Doing so must evaluate a result's effect strength, defined as a conservative assessment of practical significance. We propose the least difference in means ($\delta_L$) as a two-sample statistic that can quantify effect strength and perform a hypothesis test to determine if a result has a meaningful effect size. To facilitate consensus, $\delta_L$ allows scientists to compare effect strength between related results and choose different thresholds for hypothesis testing without recalculation. Both $\delta_L$ and the relative $\delta_L$ outperform other candidate statistics in identifying results with higher effect strength. We use real data to demonstrate how the relative $\delta_L$ compares effect strength across broadly related experiments. The relative $\delta_L$ can prioritize research based on the strength of their results.

**Main Text:**

Two-sample p-values from null hypothesis significance tests remain the gold standard for the analysis of scientific results despite calls to discontinue or de-emphasize their use (*1*, *2*). P-values can differentiate positive results (statistically significant) from null results (statistically insignificant). Yet p-values cannot give any indication of the practical significance of results: whether the observed effect size of a positive result is large enough to be considered meaningful in the real world (*3*). Practically significant results play a key role in scientific research by validating scientific hypotheses and identifying interventions that likely can provide a meaningful effect. Comparing the degree of practical significance between results also provides a means of prioritizing which interventions should be pursued with the limited resources available for scientific research. In practice, interventions with larger effect sizes have stronger practical significance and are prioritized over those with smaller effect sizes. With controlled experiments, effect size is often compared between studies with estimating the relative difference in means (DM) between populations (usually expressed as a percent change from the sample mean of a control group to an experiment group (*4*)). The relative DM allows scientists to compare effect size across a broad range of related experiments that have a combination of different treatments, measurement techniques, species, model systems, and timepoints (see Applied Examples section). Yet the relative DM disregards the uncertainty associated with the estimation. Assessing practical significance must consider the range of plausible effect sizes suggested by the sample data. Such an assessment would differentiate high effect strength results that plausibly suggest a range of all meaningful effect sizes from low effect strength results that suggest nonmeaningful effect sizes.

Evaluating the meaningfulness of the range of plausible effect sizes requires examining the data, its context, and the perspectives of scientists. Designating a result as practically significant requires some form of hypothesis test to determine if the range of plausible effect sizes are all more than a minimum threshold for what is meaningful. Selecting the value of this threshold is context-specific and can greatly differ between scientists. Scientists with differing perspectives will select different values for this threshold, yet collectively need to reach consensus for which results are practically significant. A useful statistic would allow each scientist to test for practical significance according to their threshold without having to reanalyze the data. Selecting an appropriate threshold is a critical part of this analysis and reviewing the effect strength of related results should inform this selection. A useful statistic would also facilitate threshold selection by allowing for the comparison of effect strength between results and highlight noteworthy results that have exceptionally strong practical significance. To simplify the process for data analysis, it would be ideal to use a single statistic for these tasks.

We present the least difference in means ($\delta_L$) as a statistic that is capable of all these tasks without any recalculation required. To test our statistic against other candidates, we characterize the multidimensional problem of assessing effect strength with various functions of population parameters. These functions serve as ground truth for simulation testing. We use an integrated risk assessment to test $\delta_L$ and the relative form of $\delta_L$ ($r\delta_L$) against several candidate statistics by evaluating their error rates in comparing the effect strength between simulated experiment results. Our statistics were the only candidates that demonstrated better than random error rates across all investigations. We illustrate with real data how $r\delta_L$ can be used to test for meaningful effect size and compare the effect strength of results from broadly related experiments that have a combination of different experiment models, conditions, populations, species, timepoints, treatments, and measurement techniques. We propose that reporting $r\delta_L$ of results will provide a more useful interpretation of effect size than alternative analysis techniques.

**Background**

*Bayesian Summary of Difference in Means*

Let $X_1, ..., X_m$ be an i.i.d. sample from a control group with a distribution Normal($\mu_X$, $\sigma_X^2$), and $Y_1, ..., Y_n$ be an i.i.d. sample from an experiment group with a distribution Normal($\mu_Y$, $\sigma_Y^2$). Both samples are independent from one another, and we conservatively assume unequal variance, i.e., $\sigma_X^2 \neq \sigma_Y^2$ (the Behrens-Fisher problem (*5*) for the means of normal distributions).

We analyze data in a Bayesian manner using minimal assumptions and therefore use a noninformative prior (*6*), specified as

$$p(\mu_X, \mu_Y, \sigma_X^2, \sigma_Y^2) \propto (\sigma_X^2)^{-1} (\sigma_Y^2)^{-1}. \tag{1}$$

The model has a closed-form posterior distribution. Specifically, the population means, conditional on the variance parameters and the data, follow normal distributions:

$$\mu_X | \sigma^2, x_{1:m} \sim \text{Normal}\left(\bar{x}, \frac{\sigma_X^2}{m}\right) \text{ and} \tag{2}$$

$$\mu_Y | \sigma^2, y_{1:n} \sim \text{Normal}\left(\bar{y}, \frac{\sigma_Y^2}{n}\right). \tag{3}$$

Moreover, the population variances each independently follow an inverse gamma distribution (InvGamma):

$$\sigma_X^2 | x_{1:m} \sim \text{InvGamma}\left(\frac{m-1}{2}, \frac{(m-1)s_X^2}{2}\right) \text{ and} \tag{4}$$

$$\sigma_Y^2 | y_{1:n} \sim \text{InvGamma}\left(\frac{n-1}{2}, \frac{(n-1)s_Y^2}{2}\right). \tag{5}$$

We exclude the use of prior information in this analysis because we wish to summarize the data alone and not be influenced by the beliefs of the scientist reporting the data (specifically, the strength of the prior used can considerably influence the outputs of a Bayesian statistical analysis (*7, 8*)).

*Practical Significance Over a Raw Scale*

We define that there is *stronger raw practical significance* when the absolute difference in population means ($\mu_{DM}$) is larger, where

$$|\mu_{DM}| = |\mu_Y - \mu_X|. \tag{6}$$

However, results with a positive versus a negative sign for $\mu_{DM}$ have distinct scientific interpretations and should be evaluated separately. We summarize the posterior distribution of $\mu_{DM}$ because summarizing $|\mu_{DM}|$ would disregard the sign of effect.

We first define a bounded interval that represents the lower and upper one-tailed credible bounds for $\mu_{DM}$ and encompasses all its plausible values. If $Q_{raw}(p)$ is the quantile function of this posterior at a given probability $p$, the bounds $b_{lo}$ and $b_{hi}$ satisfies

$$b_{lo} = Q_{raw}(\alpha_{DM}) \tag{7}$$

$$b_{hi} = Q_{raw}(1 - \alpha_{DM}). \tag{8}$$

However, this quantile function is difficult to compute, and there is no closed-form posterior distribution for $\mu_Y - \mu_X$ assuming unequal variances. We define $F_{raw}(x)$ as the empirical cumulative distribution function (ECDF) of $\mu_Y - \mu_X$ from $K$ Monte Carlo simulations (*9*). With $K$ samples from the posterior distribution of $\mu_X$ and $u_Y$, defined as the product of

$$\mu_X|x_{1:m} \sim t_{m-1}(\bar{x}, s_x^2/m) \text{ and} \tag{9}$$

$$\mu_y|y_{1:n} \sim t_{n-1}(\bar{y}, s_y^2/n), \tag{10}$$

the cumulative distribution function is defined as

$$F_{raw}(z) = K^{-1} \sum_{i=1}^{K} \mathbb{I}(\mu_Y^i - \mu_X^i \leq z). \tag{11}$$

We estimate the interval bounds by numerically solving for $\hat{b}_{lo}$ and $\hat{b}_{hi}$ such that.

$$F_{raw}(\hat{b}_{lo}) = \alpha_{DM}. \tag{12}$$

$$F_{raw}(\hat{b}_{hi}) = 1 - \alpha_{DM}. \tag{13}$$

We define *raw effect strength* ($S_E$) as the minimum magnitude of all values in $[\hat{b}_{lo}, \hat{b}_{hi}]$, where

$$S_E = \min_{z \in [\hat{b}_{lo}, \hat{b}_{hi}]} |z|. \tag{14}$$

Our proposed statistic, the least difference in means ($\delta_L$), is the signed version of $S_E$

$$\delta_L = \text{sign}(\bar{x}_{DM})\, S_E. \tag{15}$$

We reduce the computational complexity of this calculation by locating the minimum plausible effect size from the interval bounds rather than examining all values within the interval:

$$\delta_L = \text{sign}(\bar{x}_{DM})\, \left(\text{sign}(\hat{b}_{lo}) == \text{sign}(\hat{b}_{hi})\right)\, \min(|\hat{b}_{lo}|, |\hat{b}_{hi}|). \tag{16}$$

where $\bar{x}_{DM}$ is the difference in sample means (i.e., $\bar{y} - \bar{x}$). Summarizing the individual parts of Eq. (16), the *min* function selects the credible bound closer to zero. The equality expression sets the value of $\delta_L$ to zero if the credible bounds enclose zero (just as it does for $S_E$ in Eq. 14). Finally, $\text{sign}(\bar{x}_{DM})$ restores the original sign of the effect. Larger values of $\delta_L$ (those further from zero) convey higher effect strength between two groups' population means and suggest stronger practical significance.

Illustrations of $\delta_L$ with various credible intervals for $\mu_{DM}$ are provided in Fig 1A. We report $\delta_L$ with a percentage of $(1 - \alpha_{DM})$ to specify the credible level in the same way that credible intervals are annotated (i.e., $\mu_{DM}$ has a 95% probability of being further away from zero than the value for the 95% $\delta_L$). Colloquially, the value of $\delta_L$ represents the smallest plausible difference between the population means of the experiment group and control group supported by the data.

Note that our approach to summarizing effect strength follows the standard conventions for summarizing effect size. With the standard convention, effect size and direction are only reported when results are statistically significant (*10*) (i.e., when the confidence interval for $\mu_{DM}$ does not include zero). We similarly only report nonzero effect strength if the credible interval does not contain zero ($\delta_L \neq 0$), where we can make a Bayesian "claim with confidence" for a nonzero effect size (*10*) (we refer to this as Bayesian posterior significance (*11*)).

*Practical Significance Over a Relative Scale*

To compare the practical significance of results across loosely related experiments, we extend the concept of raw practical significance to a relative scale. We define that there is stronger *relative practical significance* when the absolute relative difference between population means ($|r\mu_{DM}|$) is larger (assuming $\mu_X > 0$), where

$$|r\mu_{DM}| = \left|\frac{\mu_Y - \mu_X}{\mu_X}\right|. \tag{17}$$

However, results with a positive versus negative sign for $r\mu_{DM}$ have distinct scientific interpretations and should be evaluated separately. We summarize the posterior distribution of $r\mu_{DM}$ because summarizing $|r\mu_{DM}|$ would disregard the sign of effect.

We first define a bounded interval that represents the lower and upper one-tailed credible bounds for $r\mu_{DM}$ and encompasses its plausible values. If $Q_{relative}(p)$ is the quantile function of this posterior at a given probability $p$, the bounds $c_{lo}$ and $c_{hi}$ satisfies

$$c_{lo} = Q_{relative}(\alpha_{DM}) \tag{18}$$

$$c_{hi} = Q_{relative}(1 - \alpha_{DM}). \tag{19}$$

However, this quantile function is difficult to compute, and there is no closed-form posterior distribution for $(\mu_Y - \mu_X)/\mu_X$. Again, we define $F_{relative}(x)$ as the empirical cumulative distribution function (ECDF) of $r\mu_{DM}$ from $K$ Monte Carlo simulations (*9*). With $K$ samples from the posterior distribution of $\mu_y$ and $u_x$, the cumulative distribution function $F_{relative}$ is defined as

$$F_{relative}(z) = K^{-1} \sum_{i=1}^{K} \mathbb{I}\left(\frac{\mu_Y^i - \mu_X^i}{\mu_X^i} \leq z\right). \tag{20}$$

We estimate the interval bounds by numerically solving for $\hat{c}_{lo}$ and $\hat{c}_{hi}$ such that

$$F_{relative}(\hat{c}_{lo}) = \alpha_{DM}, \tag{21}$$

$$F_{relative}(\hat{c}_{hi}) = 1 - \alpha_{DM}. \tag{22}$$

This interval forms a set of all plausible effect sizes for $r\mu_{DM}$. We define *relative effect strength* ($rS_E$) as the smallest effect size within this set, calculated as the minimum of the magnitude of all values within the credible interval

$$rS_E = \min_{z \in [\hat{c}_{lo}, \hat{c}_{hi}]} |z|. \tag{23}$$

Our proposed statistic, the relative least difference in means ($r\delta_L$), is the signed version of $rS_E$.

$$r\delta_L = \text{sign}(r\bar{x}_{DM})\, rS_E. \tag{24}$$

We reduce the computational complexity of this calculation by locating the minimum plausible effect size from the interval bounds rather than examining all values within the interval:

$$r\delta_L = \text{sign}(r\bar{x}_{DM}) \left(\text{sign}(\hat{c}_{lo}) == \text{sign}(\hat{c}_{hi})\right) \min(|\hat{c}_{lo}|, |\hat{c}_{hi}|). \tag{25}$$

Where $r\bar{x}_{DM}$ is the relative difference in sample means (i.e., $(\bar{y} - \bar{x}) / \bar{x}$). Summarizing the individual parts of Eq 25, the min function selects the interval bound closer to zero. The equality expression sets $r\delta_L$ to zero if the credible bounds enclose zero (just as it does for *$rS_E$*). Finally, *sign*($r\bar{x}_{DM}$) restores the original sign of the effect. Larger values of $r\delta_L$ (i.e., further from zero) convey higher relative effect strength between two groups' population means and suggest stronger relative practical significance. We report $r\delta_L$ with a percentage of (1 - $\alpha_{DM}$) to specify the credible level in the same way that credible intervals are annotated (i.e., $r\mu_{DM}$ has a 95% probability of being further away from zero than the value for the 95% $r\delta_L$). Colloquially, the value of $r\delta_L$ represents the smallest plausible relative difference between the population means of the experiment group and control group. Note that $r\delta_L$ follows the same standard statistical conventions with reporting effect size as $\delta_L$.

### *Hypothesis Testing for Meaningfulness with $\delta_L$ and $r\delta_L$*

To determine if a result is practically significant, scientists can perform a hypothesis test by testing if $\mu_{DM}$ is further from zero than a specified threshold in a particular direction. Our statistics are based on the bounds of a credible interval, which have been used for hypothesis tests against any threshold at the same credible level as the interval (*10*, *12–14*). The procedure checks if the threshold (null hypothesis) is within the bound of the interval. In this sense, intervals can be used for hypothesis testing against any threshold without recalculation of the interval. We note that there is controversy with using credible intervals for hypothesis testing because the size of the effect is estimated under the assumption that it is present (*15*). We perceive this reservation to be a nonissue because an effect size of zero has never been shown to exist in the real world and can't be confirmed with finite data (*16*). We assume a non-zero effect size is present in all cases, the question this procedure answers is whether there is evidence that it is large enough to be considered meaningful.

We approximately perform a hypothesis test that the absolute difference in means is greater than a threshold δ, specifically

$$H_0: |\mu_{DM}| \leq \delta \; ; \; H_1: |\mu_{DM}| > \delta. \tag{26}$$

However, interventions with positive signed effect sizes have different applications than those with negative sign, and the threshold should be specific to each. Results with different signed effect sizes should not be compared to each other as implied with this hypothesis test. For example, when identifying treatments for patients with high blood pressure, we are only interested in interventions that reduce blood pressure. Interventions that increase blood pressure may be useful for other applications such as patients with low blood pressure or for creating animal models of high blood pressure. As a result, the specified thresholds for positive signed and negative signed effect sizes have distinct considerations and may differ greatly.

We instead perform a composite hypothesis test that uses a positive threshold for measuring the effect strength of positive signed effect sizes and a negative threshold for negative signed effect sizes. For a positive signed effect size, we perform a hypothesis test using the threshold $\delta^+$ (where $\delta^+ > 0$) of the form

$$H_0^+: \mu_{DM} \leq \delta^+ \;;\; H_1^+: \mu_{DM} > \delta^+. \tag{27}$$

We reject $H_0^+$ and conclude practical significance if $\delta_L > \delta^+$ because $\delta_L$ is the lower bound of the posterior for $\mu_{DM}$ for positive signed effect sizes (or zero if the lower bound is less than zero).

For a negative signed effect size, we perform a hypothesis test using the threshold $\delta^-$ (where $\delta^- < 0$) of the form

$$H_0^-: \mu_{DM} \geq \delta^- \;;\; H_1^-: \mu_{DM} < \delta^- . \tag{28}$$

We reject $H_0^-$ and conclude practical significance if $\delta_L < \delta^-$ because $\delta_L$ is the upper bound of the posterior for $\mu_{DM}$ for negative signed effect sizes (or zero if the upper bound is greater than zero). Note that this composite hypothesis test does not support testing if $\mu_{DM}=0$ for the null hypothesis. The thresholds for $\delta^\pm$ must have a nonzero value because a point hypothesis is not supported for hypothesis testing within the Bayesian framework with credible intervals (*17*).

We illustrate the hypothesis testing procedure with a collection of hypothetical results from related experiments (Fig. 1B-E). For this example, two scientists choose different thresholds for meaningful positive effect size ($\delta^+$, $\delta'^+$), and negative effect size ($\delta^-$, $\delta'^-$). For typical use cases, researchers would examine positive signed effect sizes separately from negative signed, but we include both here to emphasize that they are independent examinations. Both scientists test for meaningful effect based on their own thresholds, and then arrive at a consensus for which results are practically significant.

These results are summarized by reporting the value for $\delta_L$ (Fig. 1B). Results that lack Bayesian posterior significance (credible bounds of $\mu_{DM}$ enclosing zero, where $\delta_L=0$) are excluded from any hypothesis testing because all possible hypothesis tests for meaningfulness would fail to reject $H_0$ (Fig. 1C, rows a, b). Results with positive signed effect sizes ($\delta_L>0$, Fig. 1C, rows c-f) are designated as practically significant when $\delta_L>\delta^+$ (Fig. 1C, rows d, e, f), and not practically significant when $\delta_L\leq\delta^+$ (Fig. 1C, row c). Results with negative signed effect sizes ($\delta_L<0$, rows g-j) are designated as practically significant when $\delta_L<\delta^-$ (Fig. 1C, rows i, j), and not practically significant when $\delta_L \geq \delta^-$ (Fig. 1C, rows g, h). Visually, this composite hypothesis test is equivalent to checking that the credible bounds falls outside of the null region defined by $[\delta^-, \delta^+]$ and loosely follows the procedure used in second generation p-values (*18*) when concluding full support for the alternative hypothesis (with second generation p-value equal to 1). Meanwhile, a second scientist chooses different thresholds for meaningful effect ($\delta'^-, \delta'^+$). The second scientist performs the same hypothesis testing procedure and designates practical significance for any result with a credible interval outside of the null region (Fig 2D, rows e, f, h, i, j). The scientists can reach a consensus for which results are practically significant by identifying instances where they make the same designation (Fig 2E, rows e, f, i, j). The same procedure can be performed with $r\delta_L$ using relative units for the thresholds and intervals, along with including results from more broadly related experiments (see Applied Examples section).

It is important to note that the first and second scientist may represent two scientists in the same field with differing opinions for what is meaningful or separate fields that have different requirements for the threshold. Indeed, the second scientist may even represent the first scientist in the future when their expectation for meaningful effect size is more stringent. For instance, the threshold for meaningful effect size for a biological phenomenon may be smaller in magnitude (more forgiving) for a first-in-class treatment of a particular disease in contrast to after decades of further development when many competing alternative treatments are available. The thresholds for meaningfulness are meant to change through time as competing interventions are developed.

*Measures of Raw and Relative Effect Strength*

While we have proposed two statistics to quantify the evidence of practical significance, we need to develop a structured characterization of effect strength to assess their efficacy. This assessment relies on identifying the parameters that alter effect strength on a raw and relative scale. Effect strength is difficult to characterize in a controlled fashion because it depends on several parameters in addition to $\mu_X$ and $\mu_Y$. To characterize our statistics in a controlled fashion, we decompose effect strength into a set of functions of population parameters that we have used previously to measure null strength (*19*). These functions are used as measures of effect strength just as they were used as measures of null strength. We can vary each of these in isolation and study the effects they produce on our statistics.

For assessing raw effect strength between population means, we identify a set of four measures that can be altered independently ($|\mu_{DM}|$, $\sigma_D$, $df_D$, and $\alpha_{DM}$ defined in Table 1, Figure 2 B-F, see Materials and Methods: Explanation of Raw Effect Strength Measures). For assessing the relative effect strength between population means, we divide the same effect strength measures by the control group mean when appropriate to form another set of variables that can be altered independently ($|r\mu_{DM}|$, $r\sigma_D$, $df_D$, and $\alpha_{DM}$ defined in Table 1, Figure 2 G-K, Materials and Methods: Explanation of Relative Effect Strength Measures). Note that some relative effect strength measures cannot be altered independently from raw effect strength measures (e.g., altering $|\mu_{DM}|$ can also change $|r\mu_{DM}|$ or $r\sigma_D$).

**Results**

A statistic that effectively estimates raw effect strength should covary with each measure of raw effect strength in a consistent direction. We generated a series of population parameter configurations where each measure of raw effect strength is individually altered towards higher raw effect strength (stronger evidence of raw practical significance). The mean of various candidate statistics was computed on repeated samples drawn from these configurations (candidate statistics listed in STable 1). The mean of a useful statistic could either increase for all effect strength measures or decrease. We found that only the mean values from $\delta_L$ had a significant rank correlation in a consistent direction with effect strength for all raw measures (Fig. 2L, Fig. S1-S2). Additionally, we generated sets of population parameter configurations where each measure of relative effect strength was altered towards higher relative effect strength. Only the mean values from $r\delta_L$ had a significant rank correlation in a consistent direction for all relative measures (Fig. 2M, Fig. S3-S4). However, this initial analysis had potential confounding effects since $\mu_{DM}$ and $r\mu_{DM}$ could not be altered independently from the other measures.

We next performed a risk assessment to examine how effective the candidate statistics were at determining which of two results had higher effect strength and deemed more noteworthy (based on our previous risk assessment for null strength (*19*), see Materials and Methods for explanation). We tested the comparison error, defined as the error rate associated with the candidate statistics' predictions of higher effect strength compared to the ground truth established by each effect strength measure. We averaged the frequentist risk from a collection of population parameters to assess integrated risk (*20*). Population configurations were separated based on the expected t-ratio defined as the mean t-statistic of $\mu_{DM}$ across samples scaled to the critical value (denoted as $\bar{t}_{statistic} / |t_{critical}|$, see Supplementary Materials and Methods: Parameter Space for Population Configurations). Population configurations were separated between those associated with statistical significance with positive signed effect size (expected t-ratio > 1) and negative signed effect size (expected t-ratio < -1). Investigations of comparison errors were conducted for each of the four independent measures for effect strength, both individually and simultaneously. $\delta_L$ was the only candidate statistic that exhibited an error rate lower than random 50/50 guessing for all simulation studies for raw effect strength (Fig 3A, Fig. S5-S8). Similarly, $r\delta_L$ was the only candidate statistic that exhibited an error rate lower than random for all simulation studies for relative effect strength (Fig 3B, Fig. S9-S12).

**Applied Examples**

We compiled results from studies of atherosclerosis to illustrate how the $r\delta_L$ could be used to assess the strength of practically significant results. Atherosclerosis is the underlying cause of approximately 50% of all deaths in developed nations (*21*) and is characterized by the build-up of fatty deposits, called plaques, on the inner wall of arteries. Researchers use dietary, behavioral, pharmacological, and genetic interventions to study atherosclerosis and measure various biological phenomenon to monitor disease severity, including plasma cholesterol and plaque size.

Lowering total plasma cholesterol is therapeutic in most cases (depending on the composition of the cholesterol (*21*), which is beyond the focus of this example). A multitude of studies have demonstrated results where different treatments have lowered cholesterol. Scientists must collectively decide which interventions demonstrate practical significance and have the most potential for further research, funding, commercialization, and translation. While there are many factors to consider with evaluating potential (including cost of translation, difficulty of manufacture, and likelihood of adverse side effects), effect size serves as one of the primary benchmarks. Plasma cholesterol levels vary from 60-3000 mg/dL across animal models used to research atherosclerosis and are reported in units of mmol/L as well (Table S3). This large variation in the measurement values makes it necessary to evaluate effect strength on a relative scale. The $r\delta_L$ is the only statistic that can be used to simultaneously test for meaningful effect using different thresholds and compare effect strength without recalculation (Fig. 4A-B). For this case, a review of the efficacy of existing clinical therapeutics could suggest that at least a 20% reduction in total plasma cholesterol would be a meaningful effect size. If this threshold is used to delineate a minimum meaningful effect size, these results could be separated based on a hypothesis test that designates them as practically significant ($r\delta_L < -20\%$) or not practically significant and inconclusive ($r\delta_M \geq -20\%$). Furthermore, the relative strength of the results with a meaningful effect size can be compared based on their values of $r\delta_L$, where larger values suggest stronger practical significance. It is important to note that several results that would traditionally have been associated with a strong effect size (large values of the relative difference in means) have small values of $r\delta_L$. These results (entries 5,6, and 11 of Fig. 4A) are heavily penalized because they are at the edge of statistical significance and have decreased probability of representing a reproducible result (*22*).

As a second example, a similar case study examines the practical significance of therapeutic interventions that reduce plaque size (Fig. 5A-B). Similar to measuring total cholesterol, plaque size is measured across units that span orders of magnitude (Table S4). A review of clinical results of plaque size reduction could yield a threshold of 25% for a minimum meaningful effect size. Scientists can use $r\delta_L$ to designate results as practically significant using this threshold value.

**Materials and Methods**

*Explanation of Raw Measures of Effect Strength*

Although we propose to measure effect strength of the signed posterior of $\mu_{DM}$, we do this to emphasize the importance that negative signed effect sizes should be characterized separately from positive signed effect sizes. For the sake of simplicity, we can measure effect strength as an estimate of how small $|\mu_{DM}|$ could be based on sample data (with the understanding that the two cases for effect size sign can be examined separately). From a Bayesian perspective, this can be represented with a lower quantile of a posterior distribution summarizing $|\mu_{DM}|$. Therefore, we must consider not only the location, but also the dispersion of the distribution summarizing $|\mu_{DM}|$ because both can change its lower quantiles.

Based on our definition of raw effect strength, higher effect strength is found with larger values of $|\mu_{DM}|$ with all other measures held constant, illustrated with higher effect strength from experiment 1 with its credible interval for $\mu_{DM}$ centered further from zero than experiment 2 (Fig 2A, B). Higher effect strength is also found with lower values of $\sigma_{DM}$ with all other measures held constant since the lower bound for the credible interval is closer to zero. Since $\sigma_{DM}$ is influenced by both the standard deviations and sample sizes of both groups, the contributions of each can be independently characterized with the standard deviation ($\sigma_D$) and degrees of freedom ($df_D$) of the difference between observations from group X and Y (i.e., D = Y - X).

$$\sigma_D = \sqrt{\sigma_X^2 + \sigma_Y^2} \tag{29}$$

$$df_D = m + n - 2 \tag{30}$$

There is higher effect strength with smaller values of $\sigma_D$ (contributing to $\sigma_{DM}$ in the numerator) with all other measures held constant, illustrated with higher effect strength from experiment 1 with its lower bound of the credible interval being further from zero than experiment 2 (Fig. 2C). There is also higher effect strength with larger values of $df_D$ (contributing to $\sigma_{DM}$ in the denominator) with all other measures held constant, illustrated with higher effect strength with experiment 1 from its lower bound of the credible interval further from zero (Fig. 2D). In addition to $\sigma_{DM}$ indicating how large the range of $\mu_{DM}$ could be, the specified posterior significance level ($\alpha_{DM}$) also effects the uncertainty associated with the comparison (often adjusted for experiments with multiple comparisons). There is higher effect strength with larger values of $\alpha_{DM}$ with all other measures held constant because there is an decrease in the range of possible values for $\mu_{DM}$, illustrated with higher effect strength from experiment 1 with its larger $\alpha_{DM}$ and narrower credible interval (Fig. 2E).

We have identified $|\mu_{DM}|$, $\sigma_D$, $df_D$, and $\alpha_{DM}$ as measures of effect strength (Table 1) by illustrating how changes to each of these measures in isolation leads to known changes to effect strength. Since the value of these measures can be altered independently, each of these measures can be altered as an independent measure to test the effectiveness of candidate statistics in quantifying effect strength. An effective statistic should be able to identify results with higher effect strength across all four of these measures.

### *Explanation of Relative Measures of Effect strength*

To quantify relative effect strength, we extend the measures of effect strength into units relative to the mean of the control sample. The relative difference in means ($r\mu_{DM}$) and relative standard deviation ($r\sigma_{DM}$) are normalized by the mean of the control group:

$$r\mu_{DM} = \frac{\mu_{DM}}{\mu_X} \quad (31)$$

$$r\sigma_{DM} = \frac{\sigma_{DM}}{\mu_X}. \quad (32)$$

We quantify relative effect strength by estimating the upper bound of the magnitude of $r\mu_{DM}$, where smaller values exhibit higher effect strength.

Lower relative effect strength is found with lower values of the magnitude of $r\mu_{DM}$ (abbreviated as $|r\mu_{DM}|$) with all other measures held constant, illustrated with experiment 1 having a credible interval for $r\mu_{DM}$ centered closer to zero (Fig 2G, H). Lower relative effect strength is also found with lower values $r\sigma_{DM}$ with all other measures held constant. Since $r\sigma_{DM}$ is influenced by both the relative standard deviations and sample sizes of both groups, the contributions of each can be independently characterized with the relative standard deviation ($r\sigma_D$) and degrees of freedom of the difference between observations:

$$r\sigma_D = \frac{\sigma_D}{\mu_X}. \quad (33)$$

There is lower relative effect strength with lower values of $r\sigma_D$ with all other measures held constant, illustrated with higher effect strength from experiment 1 with its narrower credible interval (Fig. 2I). There is lower relative effect strength with higher values of $df_D$ (contributing to $\sigma_{DM}$ in the denominator) with all other measures held constant, illustrated with higher effect strength from experiment 1 with its narrower credible interval (Fig. 2J). In addition to $r\sigma_{DM}$ indicating how large the range of $r\mu_{DM}$ could be, the specified posterior significance level ($\alpha_{DM}$) also effects the uncertainty associated with the comparison (often adjusted for experiments with multiple comparisons). There is higher effect strength with lower values of $\alpha_{DM}$ with all other measures held constant because there is an increase in the range of possible values for $r\mu_{DM}$, illustrated with higher effect strength from experiment 1 with its narrower credible interval (Fig

2K).

We have identified $|r\mu_{DM}|$, $r\sigma_D$, $df_D$, and $\alpha_{DM}$ as measures of relative effect strength (Table 1) by illustrating changes to each of these measures in isolation leads to known changes to relative effect strength. Since the value of these measures can be altered independently, each of these measures can be varied as independent variables to test the effectiveness of candidate statistics in quantifying relative effect strength. An effective statistic should be able to identify results with lower relative effect strength across all four of these measures.

*Integrated Risk Assessment of Effect strength*

Identifying results with higher effect strength is a critical feature for assessing practical significance. Our risk assessment is designed to benchmark the efficacy of various candidate statistics in determining which of two experiments has higher effect strength and deemed more noteworthy. In many cases, such a determination is difficult since effect strength is a function of several parameters (see Table 1). To simulate instances where it is clear which experiment has higher effect strength, we hold all population parameters constant except those that alter a specified effect strength measure (referred to as the independent measure). Using this strategy, we can then benchmark performance of the candidate statistics in determining higher effect strength for each measure of effect strength in isolation. Since the effect strength measures represent known instances where effect strength changes, we set the criterion that a successful statistic must predict higher effect strength at a rate better than random for every measure of effect strength. We use the same strategy and code that we used to quantify null strength for comparing practically insignificant results (*19*). The only change was the direction of inequalities in some of the loss functions used (see Table S2).

To accomplish this, the risk assessment must individually investigate each measure of effect strength so a direct relationship with candidate statistics' performance can be ascertained. Generating population configurations from a specific prior could not achieve these objectives.

Instead, we must carefully generate curated population configurations (See Supplementary Materials and Methods for more details).

**Discussion**

We have proposed two statistics, $\delta_L$ and $r\delta_L$, that assess the evidence of meaningful effect size by measuring effect strength. Both statistics measure effect strength by conservatively estimating the smallest effect size that is plausibly suggested by the sample data. The $\delta_L$ and $r\delta_L$ were the only candidate statistics that exhibited lower than random error in comparing effect strength across all effect strength measures. We demonstrated with applied examples how researchers can use $r\delta_L$ to assess practical significance by evaluating both the meaningfulness and effect strength of experiment results. We illustrate with real data that researchers can use $r\delta_L$ for hypothesis testing against a minimum threshold to identify practically significant results. Critically, researchers can apply different thresholds based on their differing opinions or research applications without having to re-compute $r\delta_L$: the value of the statistic remains unchanged if different thresholds are used in Fig. 4 and 5. Results that are designated as practically significant can be used to validate a scientific hypothesis or justify further research, additional funding, or translation for a particular intervention. This strategy not only favors interventions that may provide greater societal benefit, but also prioritizes a deeper scientific understanding of the most influential control mechanisms within a studied system.

Our hypothesis testing approach in Fig. 1 of examining the overlap between an interval and null region aligns closely with the procedure used to calculate second generation p-values (*16*, *18*). Indeed, both procedures would designate the same results as practically significant (i.e., full support of the alternative hypothesis) if the same threshold and intervals were used. The advantage of using $\delta_L$ is that effect strength between results can be compared regardless of their designation for practical significance, and different null regions can be used for hypothesis testing after the data is reported. In contrast to our statistic, a collection of practically significant results would all have a second-generation p-value of 0, so their effect strength cannot easily be compared (especially if different null regions are used for each result). The second-generation p-value would also require recalculation if a different null region is specified. There are similar limitations with using the Bayes Factor and two one-sided t-test p-values. We note that the treatment of these statistics in Figures 2-5 was generous because the null interval regions were fixed across simulated experiments. In practice the extent of these null regions would vary, and these candidate statistics would not be comparable across studies.

Similarly, Cohen's *d* (*23*) and related statistics (*24*) cannot be compared to thresholds for what is meaningful and between studies when effect size is standardized to the standard deviation of the sample data. Finally, interval estimation approaches (such as confidence intervals (*25*), credibility intervals, support intervals (*15*), and FDA bioequivalence procedures (*26*)) can support hypothesis testing procedures against any threshold, they cannot compare effect size clearly between results because the width and location of the interval must be compared simultaneously. We avoid this issue with $\delta_L$ by collapsing an interval estimation into a single value.

For experimental results, the presence or lack of meaningful effect size should be interpreted in the context of related results. We recommend that $r\delta_L$ should be the default statistic used to evaluate the strength of practically significant results since it allows for comparisons between a broader range of related experiments than $\delta_L$. However, $\delta_L$ would be more appropriate for cases when the control group mean is not expected to change. Reporting the $\delta_L$ or $r\delta_L$ allows scientists to better prioritize practically significant results.

**References**

1. L. G. Halsey, The reign of the p-value is over: what alternative analyses could we employ to fill the power vacuum? *Biology Letters*. **15**, 20190174 (2019).
21

**Acknowledgements**:

we would like to thank the Biocomplexity Institute at UVA for their invaluable feedback.

**Funding**: Funded by the School of Data Science, University of Virginia as part of PEB's endowment.

**Author Contributions**:

Investigation, Writing- Original draft, Visualization, Data Curation, Software: BAC.

Methodology, Formal Analysis: BAC, TRB.

Conceptualization: BAC.

Writing- Reviewing and Editing: BAC, TRB, HS, PEB.

Supervision: PEB.

**Competing Interests**: Authors declare that they have no competing interests.

**Data and Material Availability**: code used to generate all figures is written in R and available at: https://github.com/bac7wj/ACES.

**Supplementary Materials**

Supplementary Text

Fig. S1 to S12

Tables S1-S4

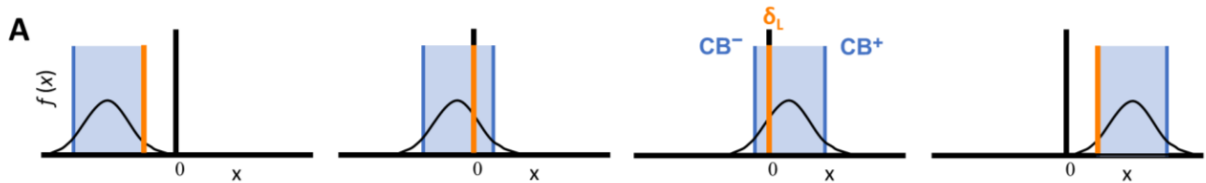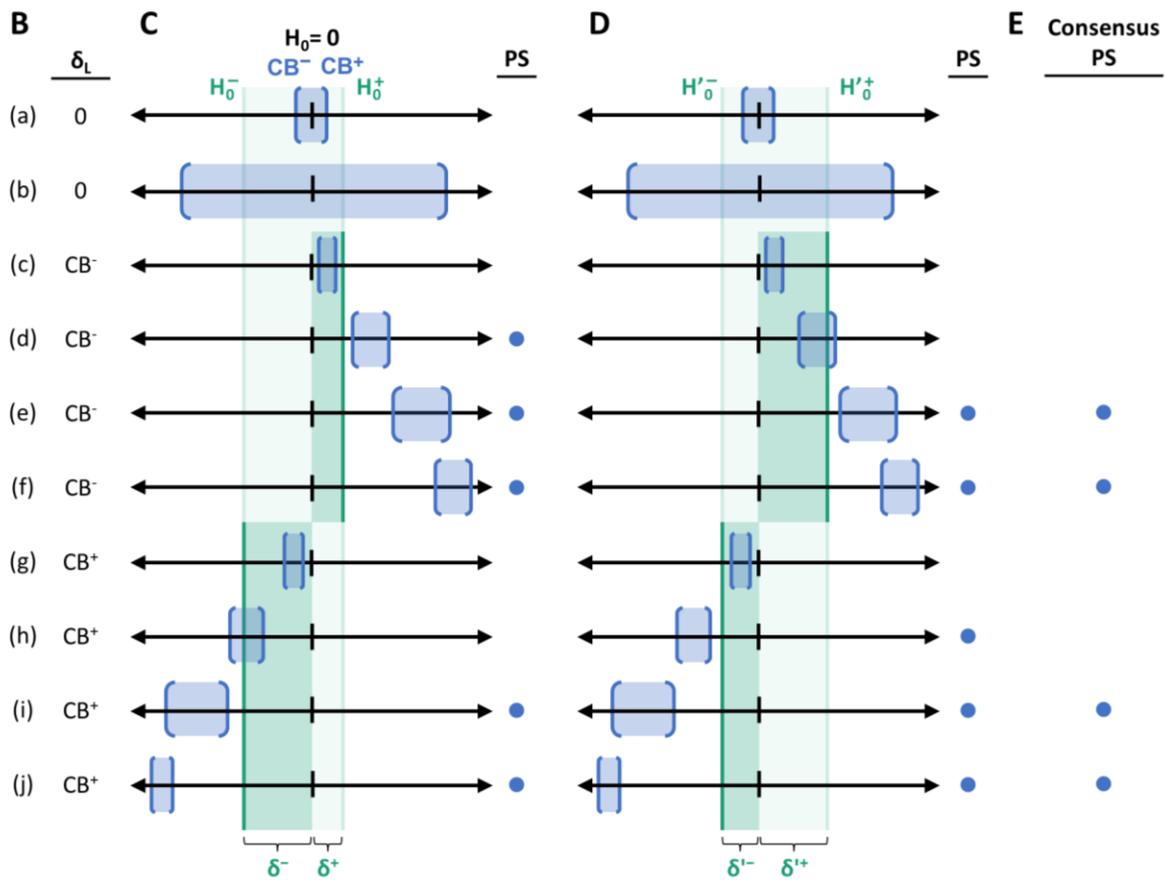

**Fig. 1: Using the least difference in means to reach consensus for practical significance.** (**A**) Several illustrations of the least difference in means statistic (orange line, $\delta_L$) as the smallest value within a credible interval (blue fill, $CB^{\pm}$) from posterior of difference in means (solid black line). (**B**) A series of hypothetical two-sample experiments with the value of $\delta_L$. (**C**) A scientist specifies a threshold for negative and positive effect sizes ($\delta^-$ and $\delta^+$, respectively) that form a null region (green interval $H_0^{\pm}$) to designate each result as practically significant (PI, entire CB interval further from zero than threshold, $\delta_L < \delta^-$ or $\delta_L > \delta^+$, blue dot) or not practically significant (CB partially or fully between threshold and zero, $\delta_L \geq \delta^-$ or $\delta_L \leq \delta^+$). (**D**) Repeated analysis can be done by a second scientist with different thresholds ($\delta'^-, \delta'^+$) with no recalculation of $\delta_L$ required. (**E**) Consensus is reached when both scientists designate a result as practically significant (blue dot).

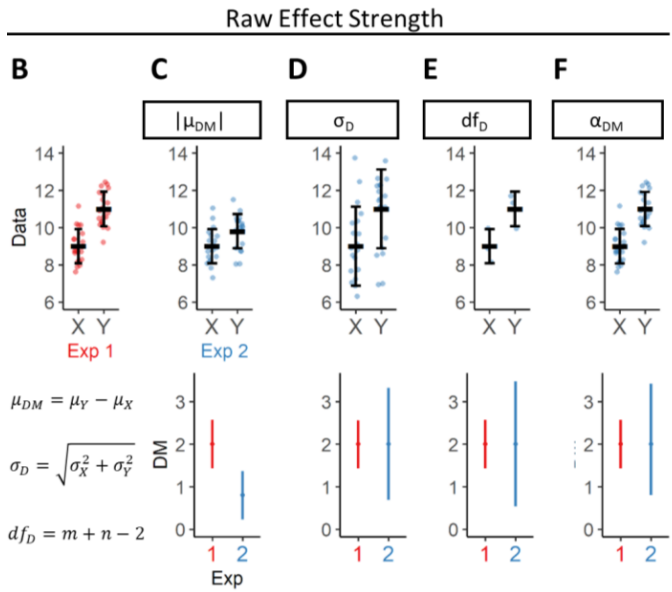
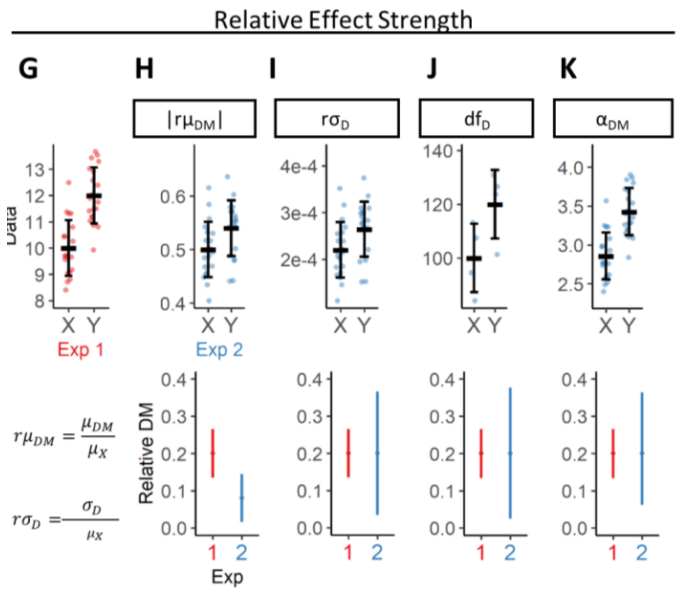

**Fig. 2: Covariation of candidate statistics with measures of effect strength.** (**A**) Data from a simulated experiment (red, Exp 1) with a control group (X) and experiment group (Y) acting as a reference to illustrate the measures of effect strength. (**B-E**) Simulated experiment data (Exp 2, blue) with lower effect strength of difference in means (DM) than Exp 1 (lower panel) via (**B**) decreased difference in means, (**C**) increased standard deviation of the difference, (**D**) decreased degrees of freedom, and (**E**) decreased posterior significance level (upper: error bars are standard

deviation, lower: error lines are 95% credible interval of the difference in means). (**F**) Heatmap of Spearman ρ of candidate statistics' mean versus each raw effect strength measure from repeated samples altered towards higher effect strength across population configurations. (**G**) Simulated data from an experiment (red, Exp 1) acting as a reference to illustrate the measures of relative effect strength. (**H-K**) Simulated experiment data (Exp 2, blue) with lower relative effect strength of relative difference in means than Exp 1 (lower panel) via (**H**) decreased relative difference in means, (**I**) increased relative standard deviation of the difference, (**J**) decreased degrees of freedom, and (**K**) decreased posterior significance level. (**L**) Heatmap of Spearman ρ of candidate statistics' mean versus each relative effect strength measure from repeated samples altered towards higher effect strength across population configurations. Asterisk denotes candidate statistic with all correlations significant and in same direction, underline denotes $p < 0.05$ for bootstrapped Spearman correlation, color displayed for significant correlations only). Abbreviations: $\bar{x}_{DM}$, $s_{DM}$, $r\bar{x}_{DM}$, $rs_{DM}$: mean, standard deviation, relative mean, and relative standard deviation of difference in sample means. $\delta_L$: least difference in means; $r\delta_L$ relative least difference in means; $\delta_M$: most difference in means; $r\delta_M$: relative most difference in means; CD: Cohen's d; $P_N$: null hypothesis testing p-value; $P_E$: two one-sided t-test equivalence p-value; $P_\delta$: second generation p-value; BF: Bayes Factor; Rnd: random 50/50 guess.

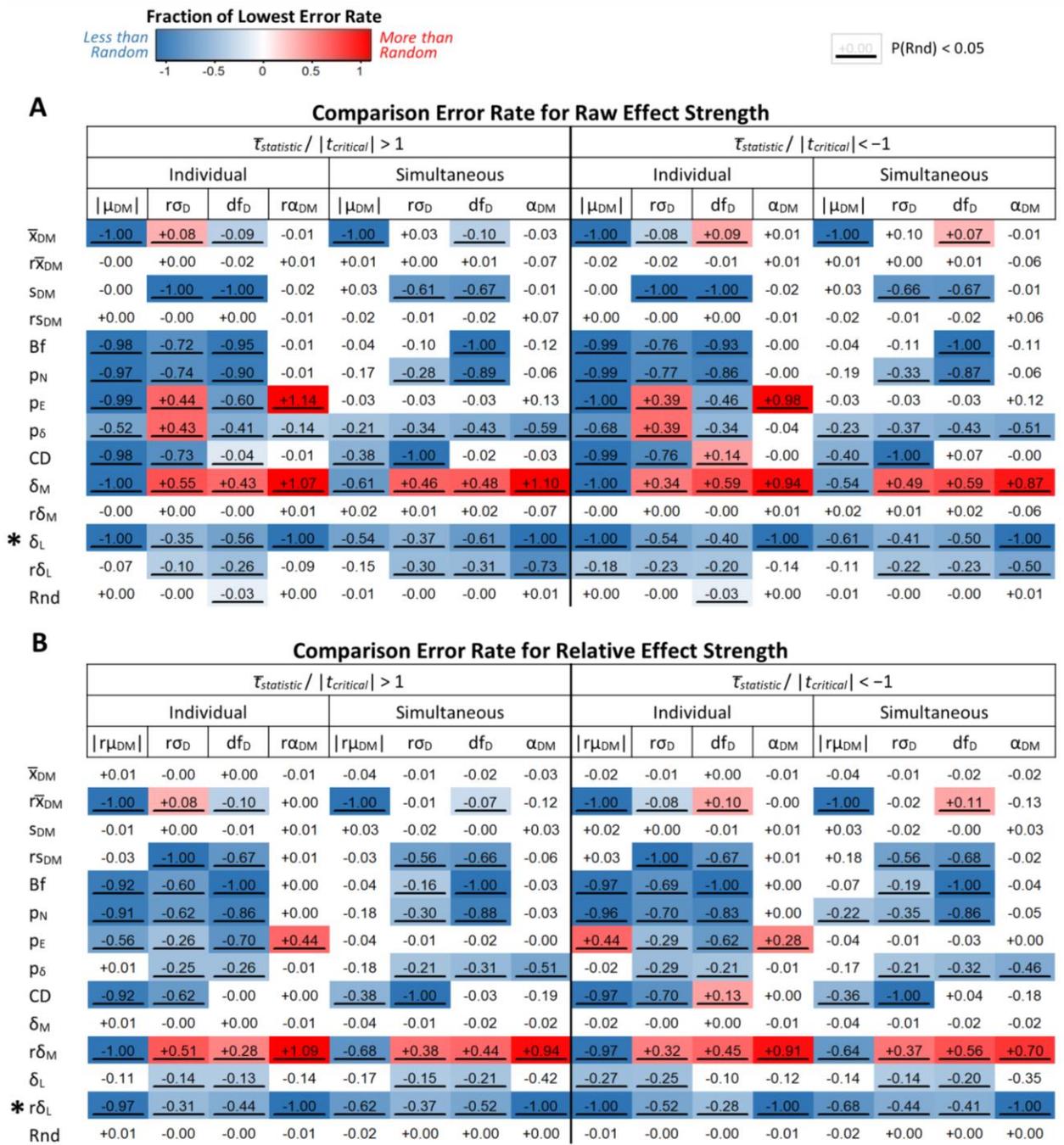

**Fig. 3: Comparison error rates of candidate statistics in identifying higher effect strength between results.** (**A**) Heatmap of comparison error rates for each candidate statistic across raw effect strength measures for identifying which of two results have higher raw effect strength. (**B**) Heatmap of comparison error rates for each candidate statistic across relative effect strength measures for identifying which of two results have higher relative effect strength. Blue fill

denotes comparison error less than random, red denotes greater than random, and white is no different than random. Numerical label in cells are comparison error rates from random behavior scaled to the lowest error rate for each column. Underlined numbers denote a comparison error rate that is statistically different than random. Investigations alter one measure of effect strength as independent variable (Individual) or several at once (Simultaneous) to serve as ground truth. Investigations are separated between population configurations associated with positive results with positive signed effect size (expected t-ratio > 1) and negative signed effect size (expected t-ratio < 1). See Fig. 2 for candidate statistic abbreviations.

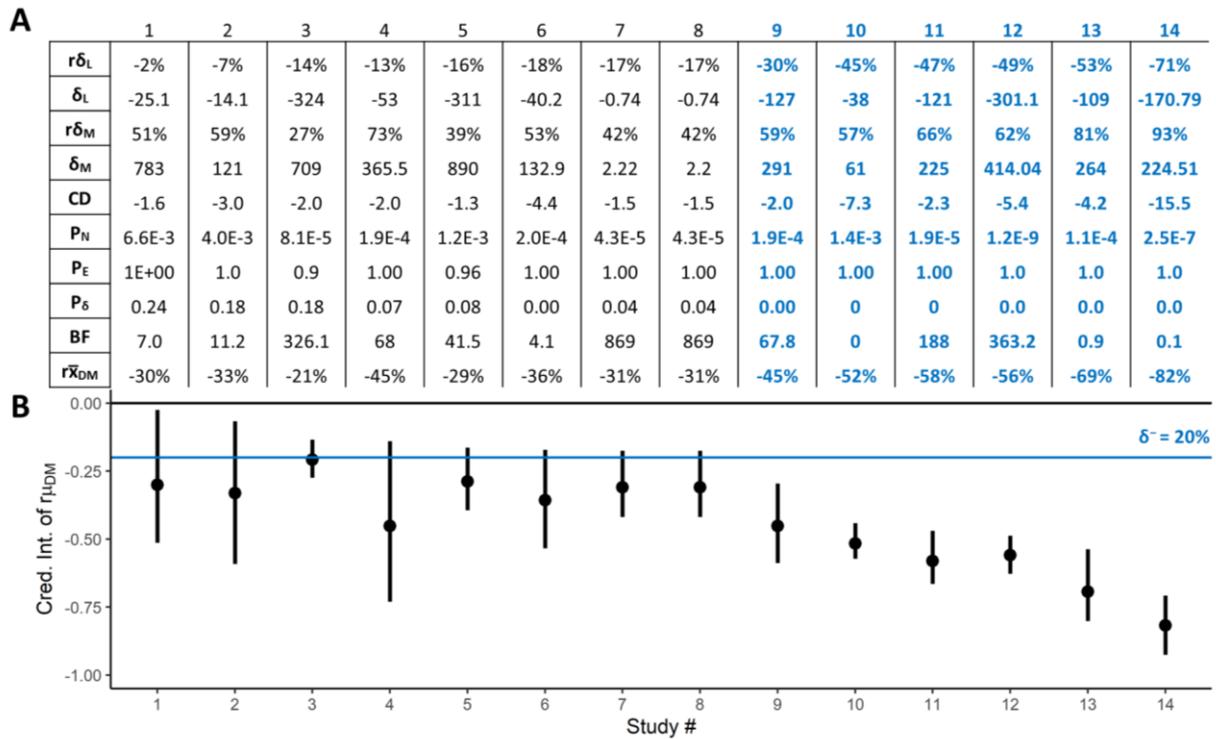

**Fig. 4: Interpreting results associated with total plasma cholesterol reduction in atherosclerosis research.** (**A**) Table of candidate statistics summarizing statistically significant results from a collection of studies (columns 1-14, practically significant results according to $r\delta_L$ highlighted in blue, $r\delta_L < \delta^-$). (**B**) For visual reference, 95% credible interval of the relative difference in means (estimated with Monte-Carlo sampling of posterior of $r\mu_{DM}$ with a noninformative uniform prior, each tail set to credible level of $(1-\alpha_{DM})$ to correspond $r\delta_L$). Null region interval was set to [-20%, +20%] of control sample mean for $P_E$, $P_\delta$, and BF. Credible intervals are Bonferroni adjusted according to each study design (see Table S3 for details and citation for each study). See Fig. 2 for candidate statistic abbreviations.

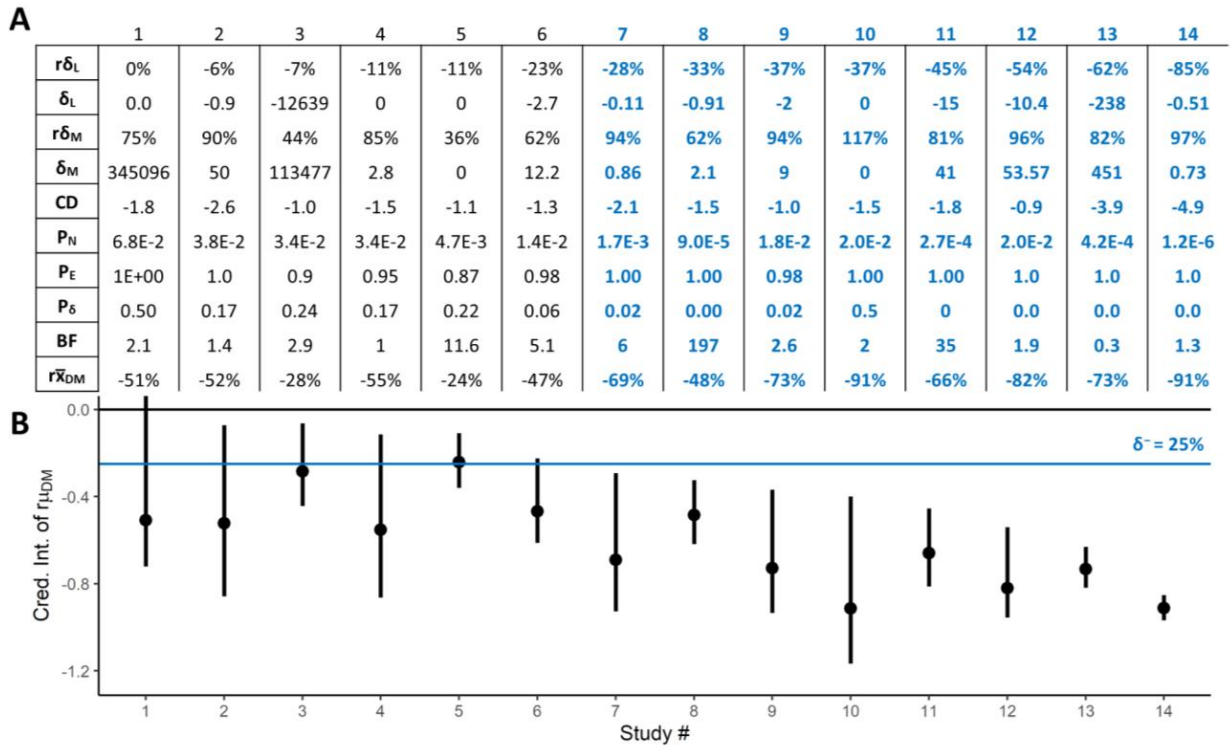

**Fig. 5: Interpreting results associated with arterial plaque size reduction in atherosclerosis research.** (**A**) Table of candidate statistics summarizing statistically significant results from a collection of studies (columns 1-14, practically significant results according to $r\delta_L$ highlighted in blue with $r\delta_L < \delta^-$). (**B**) For visual reference, 95% credible interval of the relative difference in means (estimated with Monte-Carlo sampling of posterior of $r\mu_{DM}$ with a noninformative uniform prior, each tail set to credible level of $(1-\alpha_{DM})$ to correspond $r\delta_L$). Null region interval was set to [-25%, +25%] of control sample mean for $P_E$, $P_\delta$, and BF. Credible intervals are Bonferroni adjusted according to each study design (see Table S3 for details and citation for each study). See Fig. 2 for candidate statistic abbreviations.

**Table 1: Measures of Effect Strength**

| Measure | Scale | Equation | Lower Effect Strength |
|---|---|---|---|
| $|\mu_{DM}|$ | Raw | $|\mu_{DM}| = |\mu_Y - \mu_X|,$ | + |
| $\sigma_D$ | Raw | $\sigma_D = \sqrt{\sigma_X^2 + \sigma_Y^2}$ | − |
| $df_D$ | Raw, Relative | $df_D = m + n - 2$ | + |
| $\alpha_{DM}$ | Raw, Relative |  | + |
| $|r\mu_{DM}|$ | Relative | $|r\mu_{DM}| = \left|\dfrac{\mu_{DM}}{\mu_X}\right|$ | + |
| $r\sigma_D$ | Relative | $r\sigma_D = \dfrac{\sigma_D}{\mu_X}$ | − |

*Note: Lower effect strength column indicates the direction of change for each measure to increased effect strength when other measures held constant. Abbreviations: D, difference distribution of X and Y.*

# Supplementary Materials for

## Title: The Least Difference in Means: A Statistic for Effect Size Strength and Practical Significance

Bruce A. Corliss*, Taylor R. Brown, Kevin A. Janes, Heman Shakeri, Philip E. Bourne

*Correspondence to: bac7wj@virginia.edu

**This PDF file includes:**

Supplementary Materials and Methods
Figs. S1 to S12
Tables S1 to S4
Supplement References

**Supplementary Materials and Methods**

*Literature Search*

The tables summarizing results in Fig. 4 and 5 were compiled based on a literature search using Pubmed, Google Scholar, and Google search conducted for our previous work on developing a statistic to measure practical insignificance (*1*). Included results were limited to papers that were indexed on Pubmed. Papers were identified based on searches with combinations of the following keywords:

Total cholesterol example: atherosclerosis, total cholesterol, cholesterol, plasma cholesterol, reduce, protect, increase, mouse, rabbit, human, primate, rat.

Plaque size example: plaque size, plaque area, lesion size, lesion area, reduce, protect, increase, decrease, reduce, mouse, rabbit, human, primate, rat.

The included results are not meant to be complete, but rather give the reader a simplified toy example with how the proposed statistics could be used to ascertain the practical significance of results. The mean and standard deviation of each group were either copied directly from the source publication or estimated from the figure using Web Plot Digitizer (https://automeris.io/WebPlotDigitizer/).

*Cases of Dependence Between Changes to Effect Strength Measures in Risk Assessment*

We need to test comparison error of statistics across population configurations with changes in value to each effect strength measure. In principle, we would alter the independent measure across configurations and hold all other measures constant. Yet this approach is not always possible because the raw and relative effect strength measures sometimes covary with each other. For instance, changing $|\mu_{DM}|$ across configurations must also change one or more of $\{|r\mu_{DM}|, r\sigma_{DM}, \sigma_{DM}\}$. This dependence between measures could introduce confounding relationships and prevent us from testing each measure in isolation. If confounding relationships are not dealt with, we cannot conclude if a candidate statistic can determine higher effect strength for each effect strength measure. For example, if a candidate statistic performs impressively in detecting changes to $|\mu_{DM}|$, we cannot conclude that its performance is due to changes with $|\mu_{DM}|$ since the candidate could be responding from indirect changes to $|r\mu_{DM}|$, $|r\sigma_{DM}|$, or $\sigma_{DM}$.

We avoid this confounding issue by generating sets of population configurations where the ground truth designations between the independent measure and other measures do not have any correlation. While the value of covarying measures may correlate, the ground truth designations can remain uncorrelated from each other with carefully curated population parameter datasets. To accomplish this, we varied the independent measure so that both experiments have an equal and random chance to have higher effect strength (50/50 chance). To avoid correlation with ground truth designations from other measures, we generate configurations where the other effect strength measures must either:

1) Designate experiment 1 the winner in all cases.
2) Designate experiment 2 the winner in all cases.
3) Designate experiment 1 the winner half the time, but these designations are also random and not correlated with the 50/50 designations from the independent measure.

All three cases will guarantee that there is no correlation between the ground truth designations of the independent measure and other measures. This lack of correlation is directly verified for each investigation with a binomial test ($H_0$: $\pi=0.5$) of the number of shared ground truth designations between the independent measure and each of the other measures. For an example, Fig S5A visualizes the lack of correlation of the ground truth designations for |μ$_{DM}$| compared to the other effect strength measures.

*Parameter Space for Population Configurations in Risk Assessment*

The population configurations were chosen to adequately sample the parameter space to ensure the error rates reflected general trends. We separated population configurations based on the their t-score as done previously (*1*). In brief, the t$_{critical}$ value was calculated for each population configuration and the t$_{statistic}$ was calculated for each simulated sample drawn, where

$$t_{statistic} = \frac{\bar{y} - \bar{x}}{\sqrt{\frac{s_X^2}{m} + \frac{s_Y^2}{n}}} \tag{S1}$$

$$t_{critical} := t_{\alpha_{DM},\ m+n-1}\ . \tag{S2}$$

The t-ratio is calculated by

$$t - ratio = t_{statistic}/|\bar{t}_{critical}|\ , \tag{S2}$$

Where $\bar{t}_{statistic}$ is the mean t$_{statistic}$ across all samples. Population configurations with a |t-ratio| ≤ 1 are primarily associated with null results and |t-ratio| > 1 are associated with negative signed statistically significant results (corresponding to checking if t$_{statistic}$ > |t$_{critical}$| for statistical significance). Population configurations were separated between those associated with negative signed statistically significant results with positive and negative signed effect size (t-ratio > 1 and t-ratio < -1, respectively) since they are meant to be analyzed separately. Population configurations associated with null results were not tested because effect strength is not designed to be comparable for statistically insignificant results.

*Simultaneous Risk Assessment in Risk Assessment*

Our approach of varying a single measure of effect strength at a time for the risk assessment for effect strength unfortunately does not simulate real world conditions. It would be reasonable to expect multiple measures of effect strength to vary simultaneously when comparing effect strength between results. To address this shortcoming, we designed population configurations that had multiple effect strength measures varied simultaneously as a more realistic scenario.

We designed a set of population configurations that allowed for all four raw effect strength measures to vary simultaneously (Fig 3A, columns under "Simultaneous" header). We examined whether candidate statistics could predict effect strength in a better than random fashion by comparing the predicted designations to the ground truth designations for each raw effect strength measure. Another set of population configurations were generated that allowed for all relative effect strength measures to change simultaneously (Fig 3B, columns under "Simultaneous" header). We examined whether candidate statistics could predict effect strength in a better than random fashion by comparing the predicted designations to the ground truth designations for each relative effect strength measure.

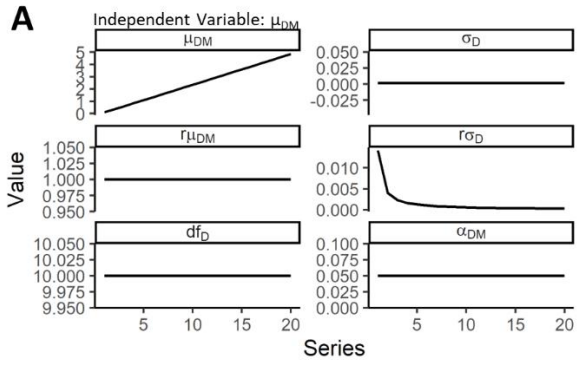
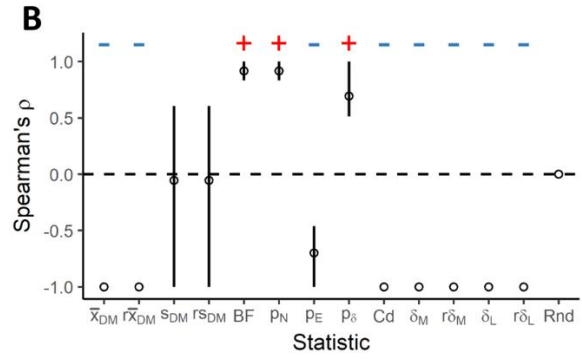
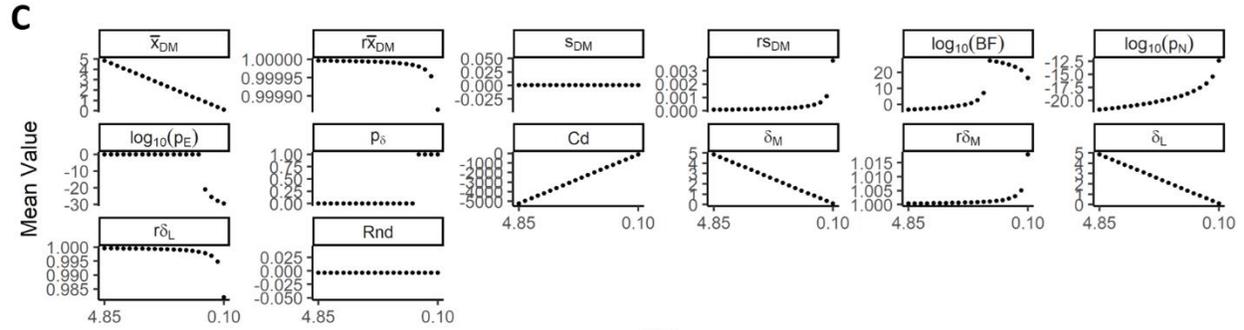
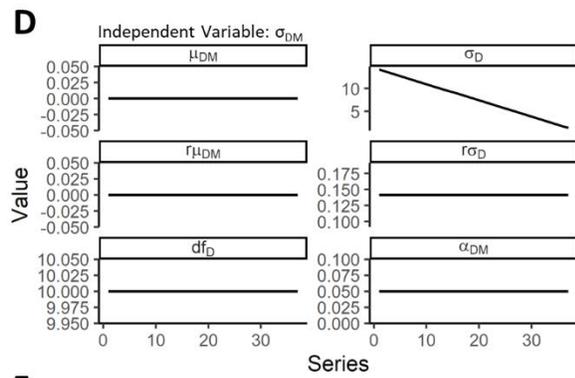
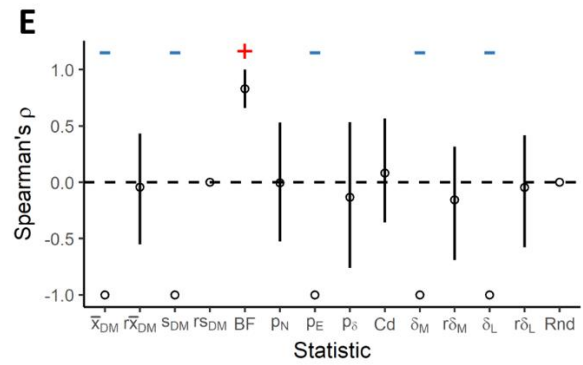
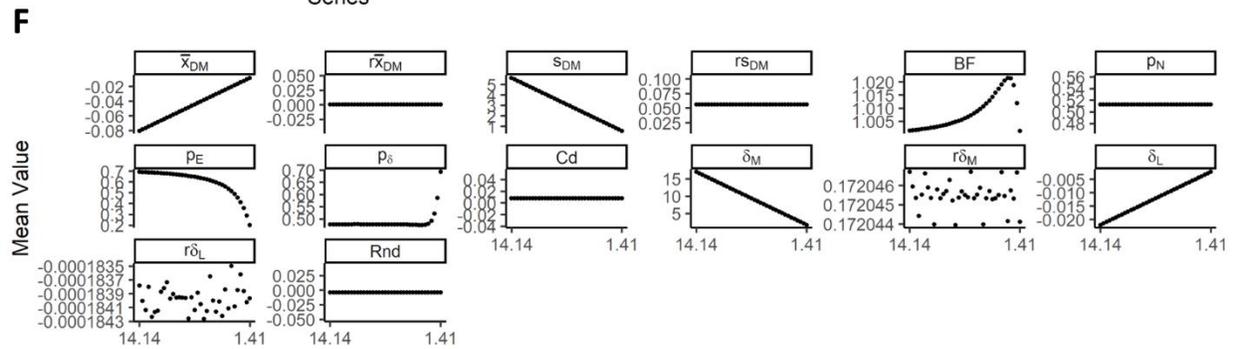

**Fig. S1: Correlation of candidate statistics versus µ$_{DM}$ and σ$_D$ towards stronger raw practical significance.** (**A**) A series of population configurations with decreasing µ$_{DM}$ towards higher effect strength (changes to µ$_{DM}$ could not be completely isolated from all other effect strength measures, so rσ$_D$ also changed with this series). (**B**) Spearman's ρ of mean of each candidate statistic versus µ$_{DM}$ and (**C**) mean value of candidate statistic across configurations. (**D**) A series of population configurations with decreasing σ$_D$ towards higher effect strength with all other effect strength measures held constant. (**E**) Spearman's ρ of mean of each candidate statistic versus σ$_D$ and (**F**) mean value of candidate statistic across configurations. (B, E) Error bars are 95% confidence interval of Spearman's ρ with Bonferroni correction, with red plus denoting candidate statistics with a significant positive correlation and blue minus denoting a significance negative correlation (1E3 samples drawn from each population configuration in the series).

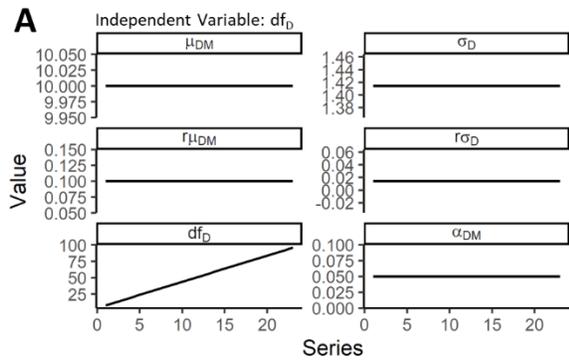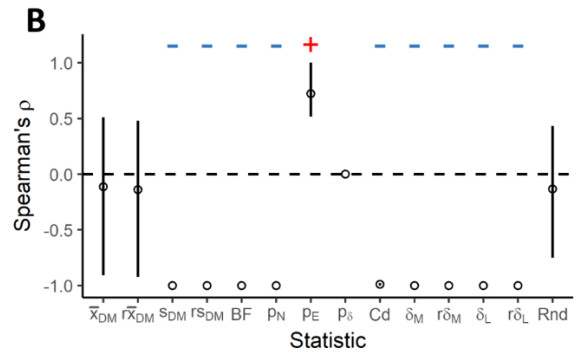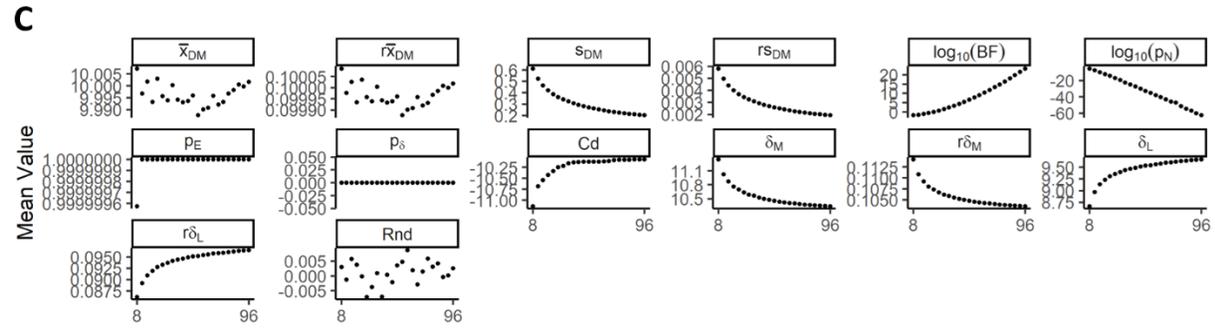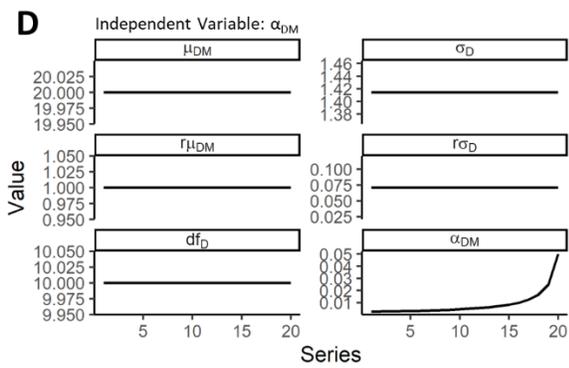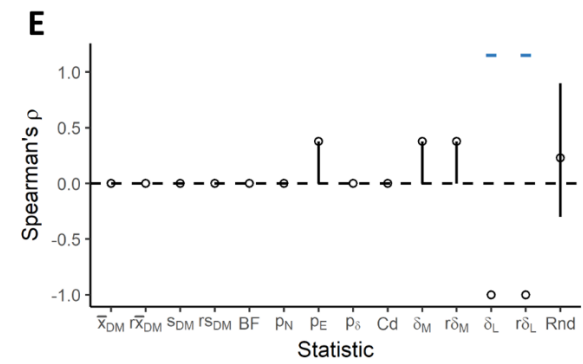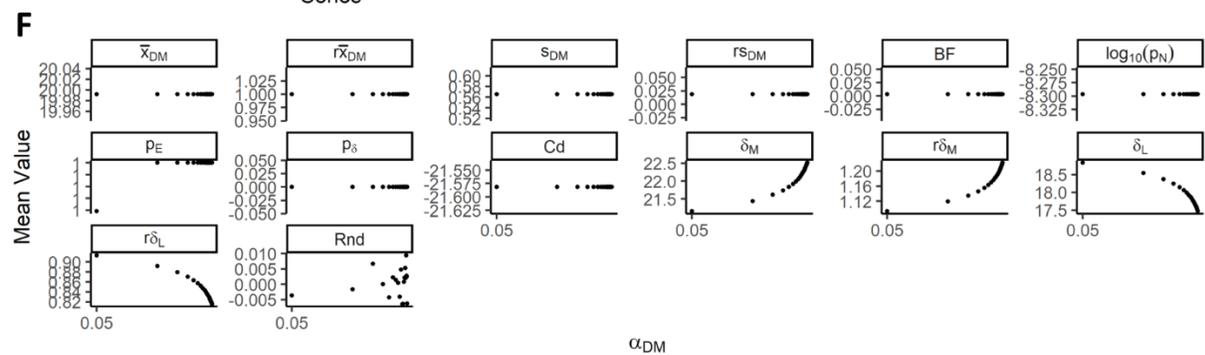

**Fig. S2: Correlation of candidate statistics versus df$_D$ and α$_{DM}$ towards stronger raw practical significance.** (**A**) A series of population configurations with decreasing df$_D$ toward higher effect strength with all other effect strength measures held constant. (**B**) Spearman's ρ of mean of each candidate statistic versus df$_D$ and (**C**) mean value of candidate statistic across configurations. (**D**) A series of population configurations with decreasing α$_{DM}$ toward higher effect strength with all other effect strength measures held constant. (**E**) Spearman's ρ of mean of each candidate statistic versus α$_{DM}$ and (**F**) mean value of candidate statistic across configurations. (B, E) Error bars are 95% confidence interval of Spearman's ρ with Bonferroni correction, with red plus denoting candidate statistics with a significant positive correlation and blue minus denoting a significance negative correlation (1E3 samples drawn from each population configuration in the series).

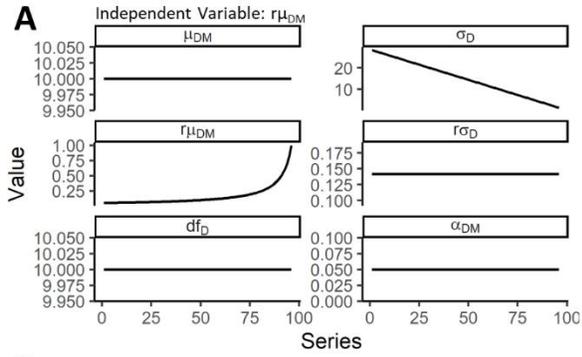
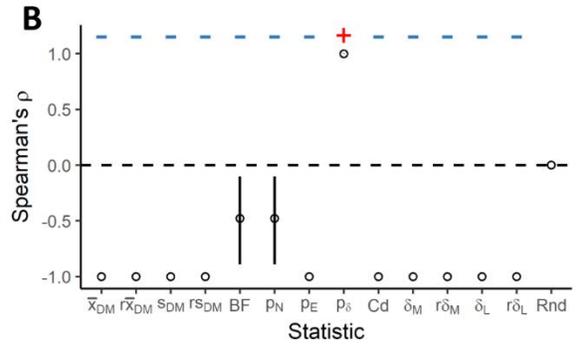
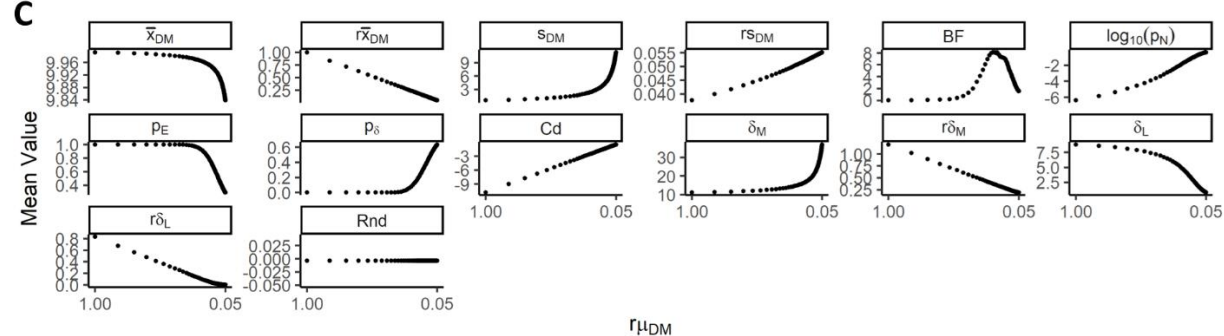
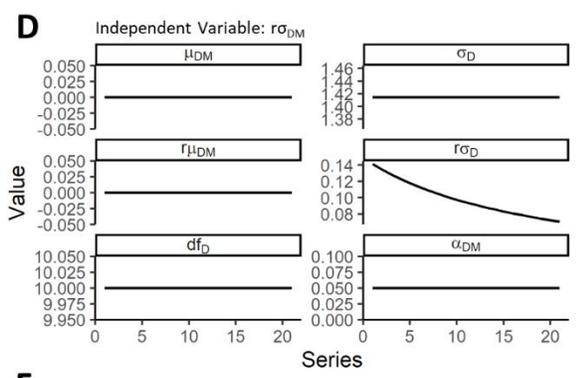
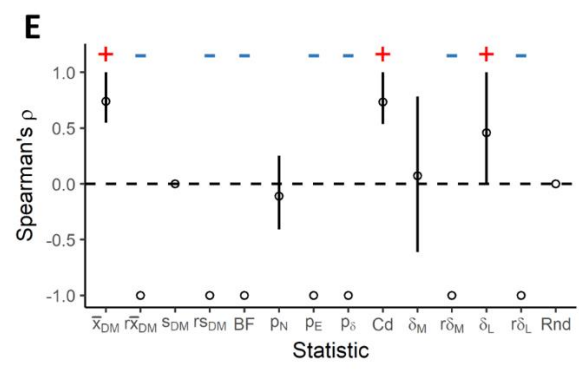
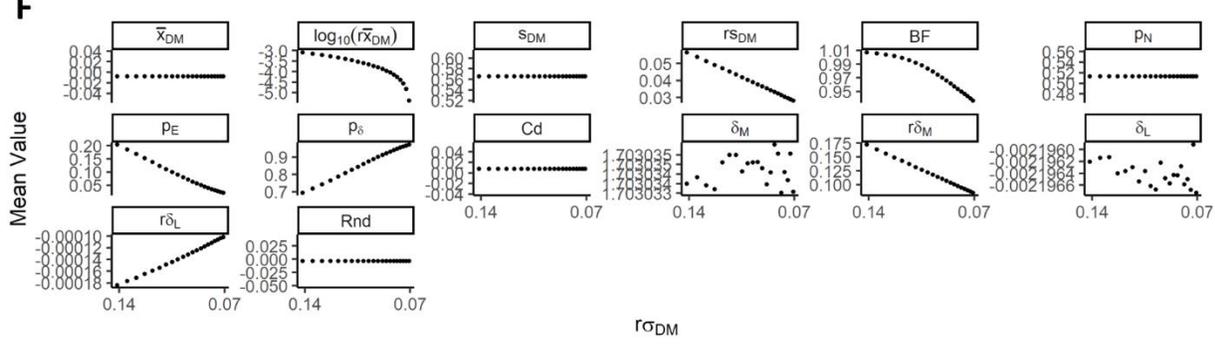

**Fig. S3: Correlation of candidate statistics versus r$\mu_{DM}$ and r$\sigma_D$ towards stronger relative practical significance.** (**A**) A series of population configurations with decreasing r$\mu_{DM}$ towards stronger relative effect strength (changes to r$\mu_{DM}$ could not be completely isolated from all variables, so $\sigma_D$ also changed with this series). (**B**) Spearman's $\rho$ of mean of each candidate statistic versus r$\mu_{DM}$ and (**C**) mean value of candidate statistic across configurations. (**D**) A series of population configurations with decreasing r$\sigma_D$ towards higher relative effect strength with all other effect strength measures held constant. (**E**) Spearman's $\rho$ of mean of each candidate statistic versus r$\sigma_D$ and (**F**) mean value of candidate statistic across configurations. (B, E) Error bars are 95% confidence interval of Spearman's $\rho$ with Bonferroni correction, with red plus denoting candidate statistics with a significant positive correlation and blue minus denoting a significance negative correlation (1E3 samples drawn from each population configuration in the series).

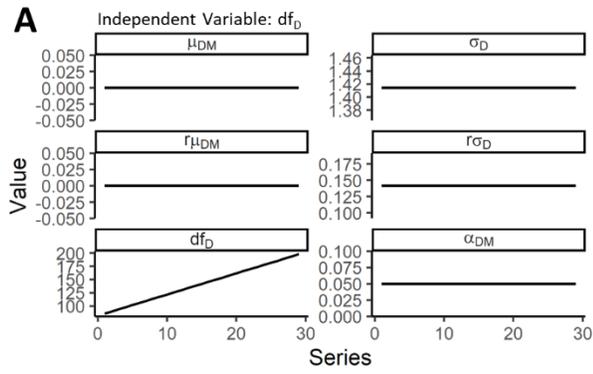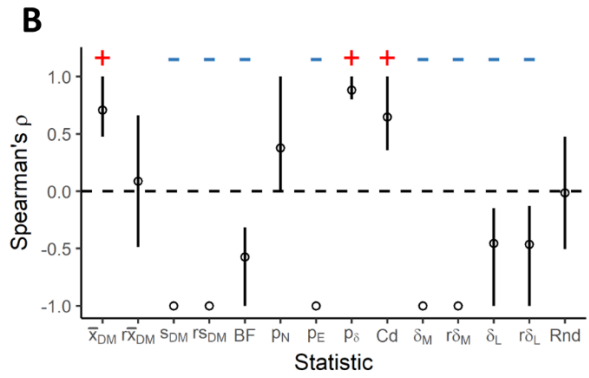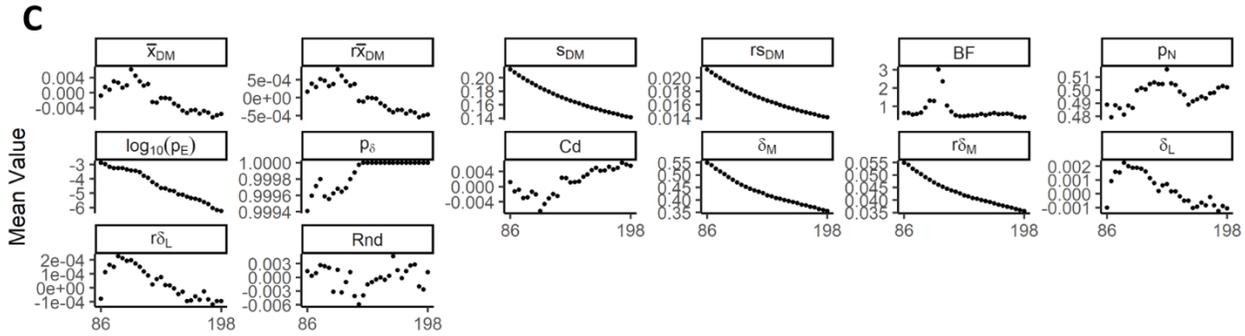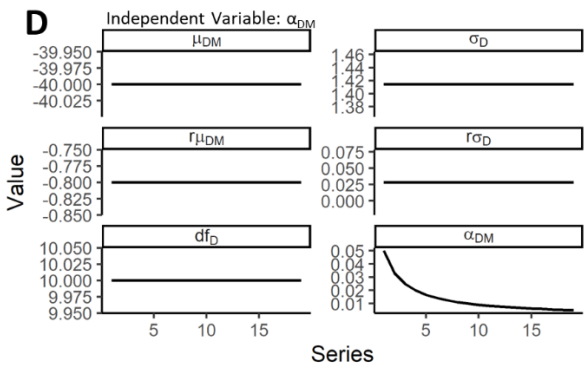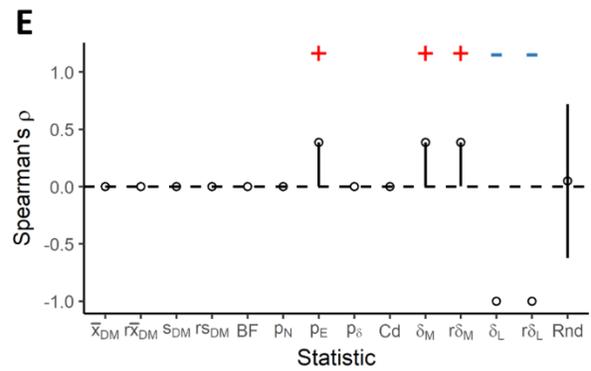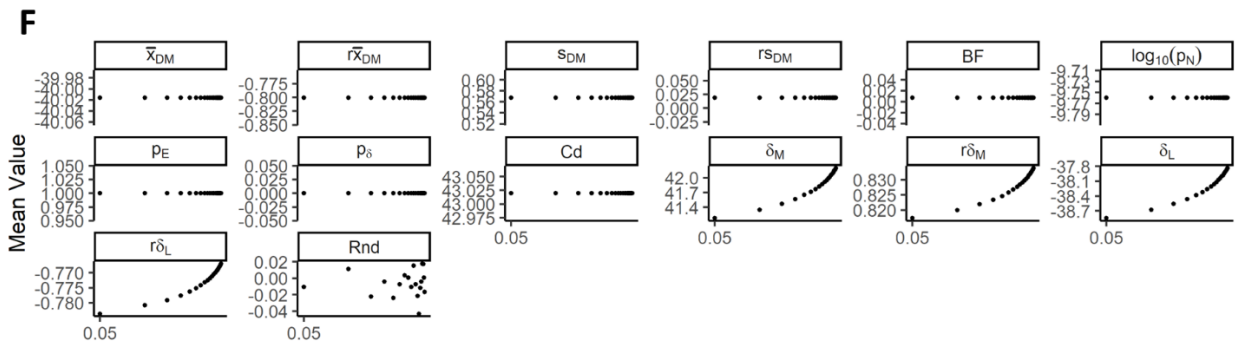

**Fig. S4: Correlation of candidate statistics versus $df_D$ and $α_{DM}$ towards stronger relative practical significance.** (**A**) A series of population configurations with decreasing $df_D$ toward higher relative effect strength with all other effect strength measures held constant. (**B**) Spearman's ρ of mean of each candidate statistic versus $df_D$ and (**C**) mean value of candidate statistic across configurations. (**D**) A series of population configurations with $α_{DM}$ reduced toward higher relative effect strength with all other effect strength measures held constant. (**E**) Spearman's ρ of mean of each candidate statistic versus $α_{DM}$ and (**F**) mean value of candidate statistic across configurations. (B, E) Error bars are 95% confidence interval of Spearman's ρ with Bonferroni correction, with red plus denoting candidate statistics with a significant positive correlation and blue minus denoting a significant negative correlation (1E3 samples drawn from each population configuration in the series).

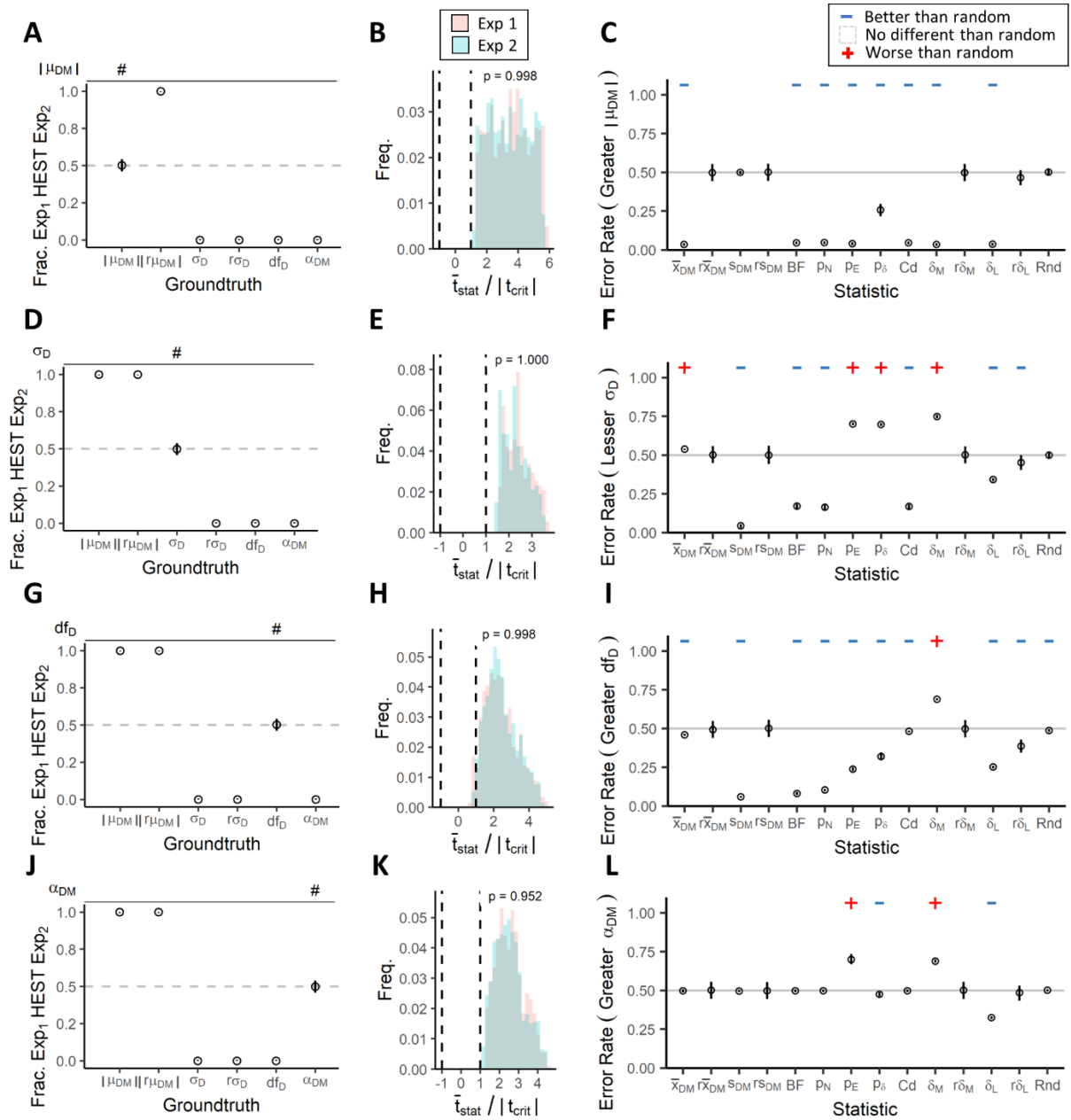

**Fig. S5: $\delta_L$ is the only statistic that has lower than random comparison error with positive signed results for each effect strength measure.** (**A**) Fraction of population configurations where experiment 1 has higher effect strength than (HEST) experiment 2 according to each effect strength measure, with $\mu_{DM}$ serving as ground truth. (**B**) Histogram of the t-ratio for population configurations, indicating that results from experiment 1 (blue) and experiment 2 (pink) are both associated with positive signed statistically significant results ($\bar{t}_{statistic} / |t_{critical}| > 1$). (**C**) Mean comparison error rate of candidate statistics in identifying which experiment has higher effect strength via higher $\mu_{DM}$ across population configurations (50 observations per sample). (**D**) Fraction of population configurations where experiment 1 has higher effect strength than experiment 2 according to each effect strength measure, with $\sigma_D$ serving as ground truth. (**E**) Histogram of the t-ratio indicating population configurations are associated with positive signed statically significant results. (**F**) Mean comparison error rate of candidate statistics in identifying which experiment has higher effect strength via lower $\sigma_D$ across population configurations (50 observations per sample). (**G**) Fraction of population configurations where experiment 1 has higher effect strength than experiment 2 according to each effect strength measure, with $df_D$ serving as ground truth. (**H**) Histogram of the t-ratio indicating population configurations are associated with positive signed statistically significant results. (**I**) Mean comparison error rate of candidate statistics in identifying which experiment has higher effect strength via higher $df_D$ across population configurations (6 - 40 observations per sample). (**J**) Fraction of population configurations where experiment 1 has higher effect strength than experiment 2 according to each effect strength measure, with $\alpha_{DM}$ serving as ground truth. (**K**) Histogram of the t-ratio indicating population configurations are associated with positive signed statistically significant results. (**L**) Mean comparison error rate of candidate statistics in identifying which experiment has higher effect strength via higher $\alpha_{DM}$ across population configurations (30 observations per sample). (A, D, G, J) '#' denotes measures that have a nonrandom number of shared designations with independent measures (listed at top of y-axis) for which experiments are designated with higher effect strength ($p < 0.05$ from Bonferroni corrected two-tailed binomial test for coefficient equal to 0.5 between independent measure and each effect strength measure). (B, E, H, K) Discrete Kolmogorov-Smirnov test between histograms. (C, F, I, L) Pairwise t-test with Bonferroni correction for all combinations, where blue minus denotes a mean error rate lower than random, red plus denotes higher than random. N=1E3 population configurations generated for each study, n=1E2 samples drawn per configuration.

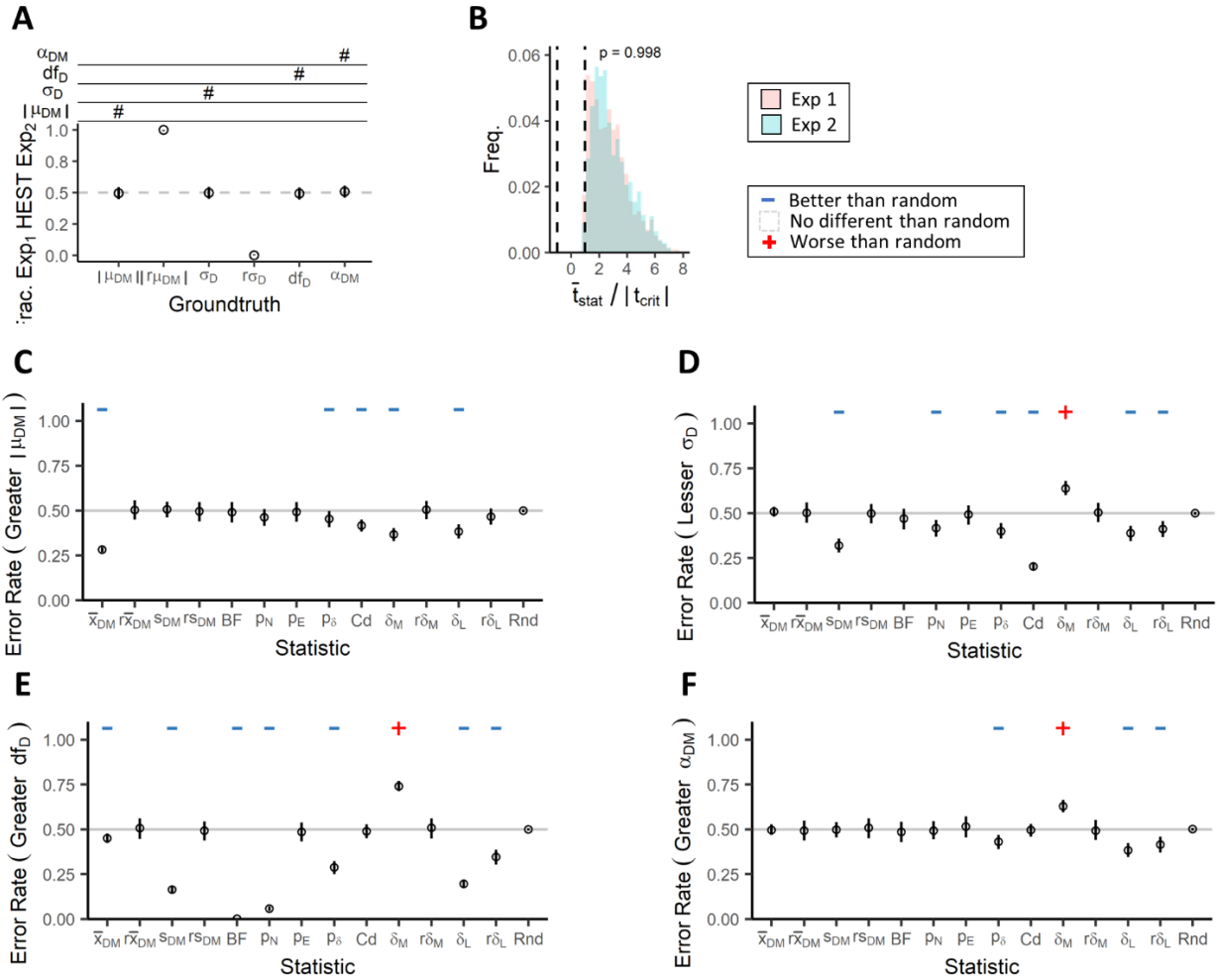

**Fig. S6: $\delta_L$ is the only statistic that has lower than random comparison error with positive signed statistically significant results across all effect strength measures simultaneously.** (**A**) Fraction of population configurations where experiment 1 has higher effect strength than (HEST) experiment 2 according to each effect strength measure, with $\mu_{DM}$, $\sigma_D$, $df_D$, and $\alpha_{DM}$ serving as separate ground truths simultaneously. (**B**) Histogram of the t-ratio for population configurations, indicating that results from experiment 1 (blue) and experiment 2 (pink) are both associated with positive signed statistically significant results ($\bar{t}_{statistic} / |t_{critical}| > 1$). From a single data set, mean comparison error rate of candidate statistics in identifying which experiment has higher effect strength via (**C**) higher $\mu_{DM}$, (**D**) lower $\sigma_D$, (**E**) higher $df_D$, and (**F**) higher $\alpha_{DM}$ across population configurations. (A) '#' denotes measures that have a nonrandom number of shared designations with each independent measure (listed at top of y-axis) for which experiments are designated with higher effect strength ($p < 0.05$ from Bonferroni corrected two-tailed binomial test for coefficient equal to 0.5 between each independent measure and every effect strength measure). (B) Discrete Kolmogorov-Smirnov test between histograms. (C, D, E, F) Pairwise t-test with Bonferroni correction for all combinations, where blue minus denotes a mean error rate lower than random, red plus denotes higher than random. N=1E3 population configurations generated for each study, n=1E2 samples drawn per configuration, 5 - 20 observations per sample.

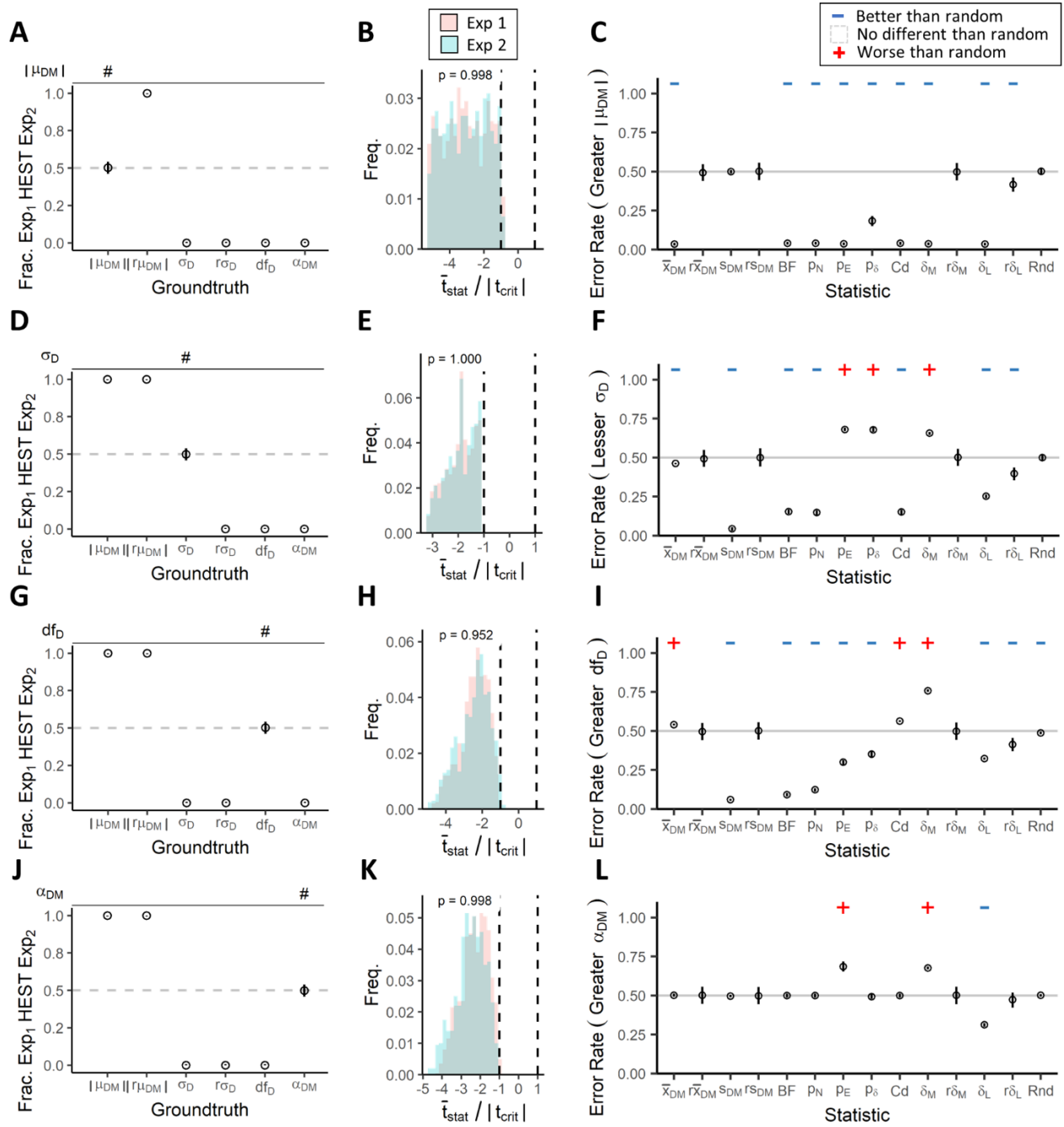

**Fig. S7: $\delta_L$ is the only statistic that has lower than random comparison error with negative signed statistically significant results for each effect strength measure.** (**A**) Fraction of population configurations where experiment 1 has higher effect strength than (HEST) experiment 2 according to each effect strength measure, with $\mu_{DM}$ serving as ground truth. (**B**) Histogram of the t-ratio for population configurations, indicating that results from experiment 1 (blue) and experiment 2 (pink) are both associated with negative signed statistically significant results ($\bar{t}_{statistic} / |t_{critical}| < -1$). (**C**) Mean comparison error rate of candidate statistics in identifying which experiment has higher effect strength via higher $\mu_{DM}$ across population configurations (50 observations per sample). (**D**) Fraction of population configurations where experiment 1 has higher effect strength than experiment 2 according to each effect strength measure, with $\sigma_D$ serving as ground truth. (**E**) Histogram of the t-ratio indicating that population configurations are associated with negative signed statistically significant results. (**F**) Mean comparison error rate of candidate statistics in identifying which experiment has higher effect strength via lower $\sigma_D$ across population configurations (50 observations per sample). (**G**) Fraction of population configurations where experiment 1 has higher effect strength than experiment 2 according to each effect strength measure, with $df_D$ serving as ground truth. (**H**) Histogram of the t-ratio indicating that population configurations are associated with negative signed statistically significant results. (**I**) Mean comparison error rate of candidate statistics in identifying which experiment has higher effect strength via higher $df_D$ across population configurations (6 - 40 observations per sample). (**J**) Fraction of population configurations where experiment 1 has higher effect strength than experiment 2 according to each effect strength measure, with $\alpha_{DM}$ serving as ground truth. (**K**) Histogram of the t-ratio indicating that population configurations are associated with negative signed statistically significant results. (**L**) Mean comparison error rate of candidate statistics in identifying which experiment has higher effect strength via higher $\alpha_{DM}$ across population configurations (30 observations per sample). (A, D, G, J) '#' denotes measures that have a nonrandom number of shared designations with independent measure (listed at top of y-axis) for which experiments are designated with higher effect strength (p < 0.05 from Bonferroni corrected two-tailed binomial test for coefficient equal to 0.5 between independent measure and each effect strength measure). (B, E, H, K) Discrete Kolmogorov-Smirnov test between histograms. (C, F, I, L) Pairwise t-test with Bonferroni correction for all combinations, where blue minus denotes a mean error rate lower than random, red plus denotes higher than random. N=1E3 population configurations generated for each study, n=1E2 samples drawn per configuration.

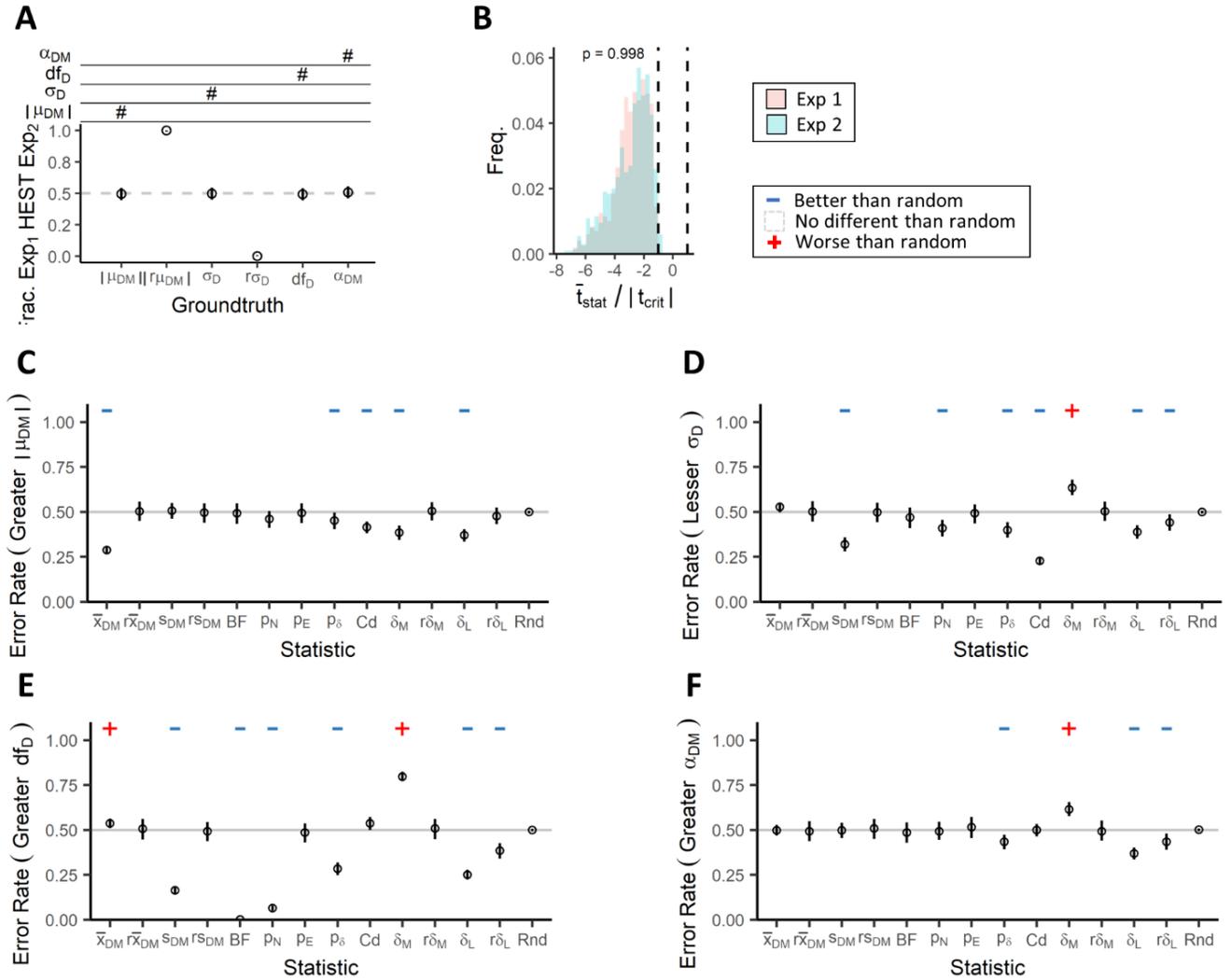

**Fig. S8: $\delta_L$ is the only statistic that has lower than random comparison error with negative signed statistically significant results across all effect strength measures simultaneously.** (**A**) Fraction of population configurations where experiment 1 has higher effect strength than (HEST) experiment 2 according to each effect strength measure, with $\mu_{DM}$, $\sigma_D$, $df_D$, and $\alpha_{DM}$ serving as separate ground truths simultaneously. (**B**) Histogram of the t-ratio for population configurations, indicating that results from experiment 1 (blue) and experiment 2 (pink) are both associated with negative signed statistically significant results ($\bar{t}_{statistic} / |t_{critical}| < -1$). From a single data set, mean comparison error rate of candidate statistics in identifying which experiment has higher effect strength via (**C**) higher $\mu_{DM}$, (**D**) lower $\sigma_D$, (**E**) higher $df_D$, and (**F**) higher $\alpha_{DM}$ across population configurations. (A) '#' denotes measures that have a nonrandom number of shared designations with each independent measure (listed at top of y-axis) for which experiments are designated with higher effect strength ($p < 0.05$ from Bonferroni corrected two-tailed binomial test for coefficient equal to 0.5 between each independent measure and every effect strength measure). (B) Discrete Kolmogorov-Smirnov test between histograms. (C, D, E, F) Pairwise t-test with Bonferroni correction for all combinations, where blue minus denotes a mean error rate lower than random, red plus denotes higher than random. N=1E3 population configurations generated for each study, n=1E2 samples drawn per configuration, 6 - 30 observations per sample.

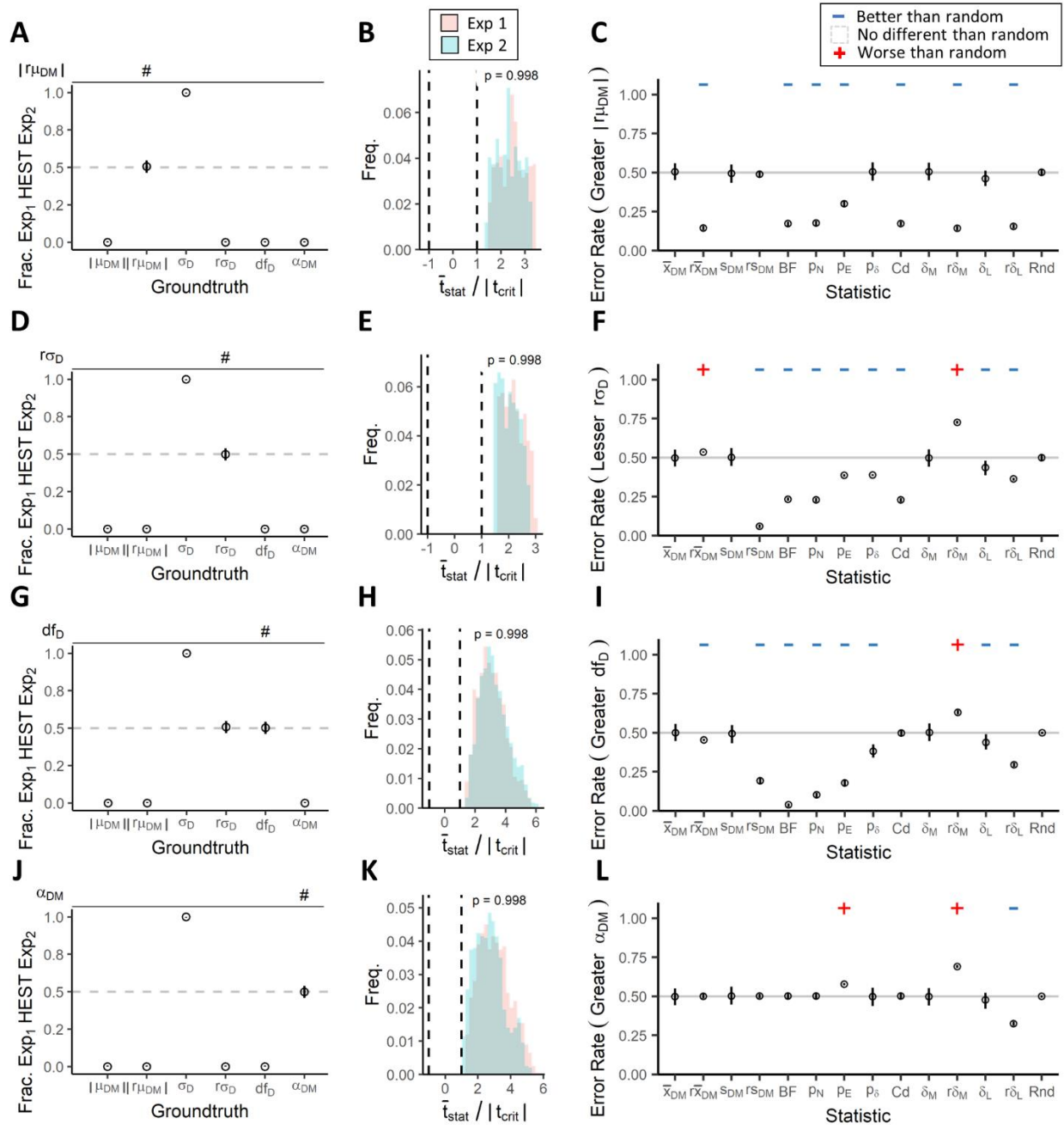

**Fig. S9: $r\delta_L$ is the only statistic that has lower than random comparison error for positive signed statistically significant results for each measure of relative effect strength.** (**A**) Fraction of population configurations where experiment 1 has higher effect strength than (HEST) experiment 2 according to each effect strength measure, with $r\mu_{DM}$ serving as ground truth. (**B**) Histogram of the t-ratio for population configurations, indicating that results from experiment 1 (blue) and experiment 2 (pink) are both associated with positive signed statistically significant results ($\bar{t}_{statistic} / |t_{critical}| > 1$). (**C**) Mean comparison error rate of candidate statistics in identifying which experiment has higher relative effect strength via higher $r\mu_{DM}$ across population configurations (50 observations per sample). (**D**) Fraction of population configurations where experiment 1 has higher effect strength than experiment 2 according to each effect strength measure, with $r\sigma_D$ serving as ground truth. (**E**) Histogram of the t-ratio indicating population configurations are associated with positive signed statistically significant results. (**F**) Mean comparison error rate of candidate statistics in identifying which experiment has higher relative effect strength via lower $r\sigma_D$ across population configurations (50 observations per sample). (**G**) Fraction of population configurations where experiment 1 has higher effect strength than experiment 2 according to each effect strength measure, with $df_D$ serving as ground truth. (**H**) Histogram of the t-ratio indicating population configurations are associated with positive signed statistically significant results. (**I**) Mean comparison error rate of candidate statistics in identifying which experiment has higher relative effect strength via higher $df_D$ across population configurations (6 - 30 observations per sample). (**J**) Fraction of population configurations where experiment 1 has higher effect strength than experiment 2 according to each effect strength measure, with $\alpha_{DM}$ serving as ground truth. (**K**) Histogram of the t-ratio indicating population configurations are associated with positive signed statistically significant results. (**L**) Mean comparison error rate of candidate statistics in identifying which experiment has higher relative effect strength via higher $\alpha_{DM}$ across population configurations (30 observations per sample). (A, D, G, J) '#' denotes measures that have a nonrandom number of shared designations with independent measure (listed at top of y-axis) for which experiments are designated with higher effect strength ($p < 0.05$ from Bonferroni corrected two-tailed binomial test for coefficient equal to 0.5 between independent measure and each effect strength measure). (B, E, H, K) Discrete Kolmogorov-Smirnov test between histograms. (C, F, I, L) Pairwise t-test with Bonferroni correction for all combinations, where blue minus denotes a mean error rate lower than random, red plus denotes higher than random. N=1E3 population configurations generated for each study, n=1E2 samples drawn per configuration.

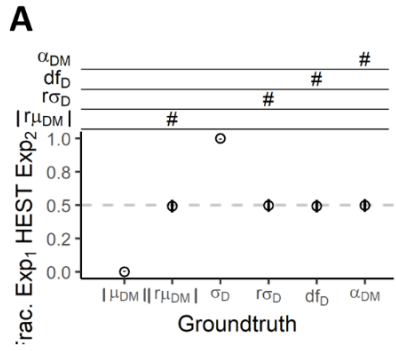
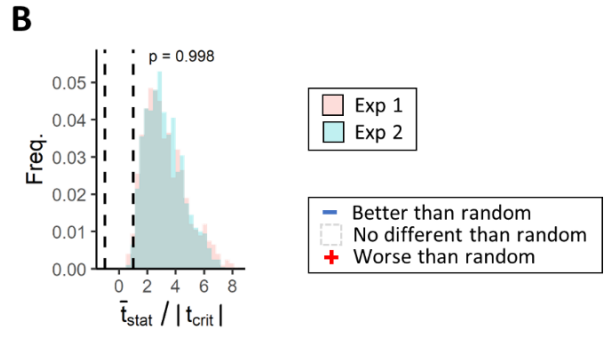
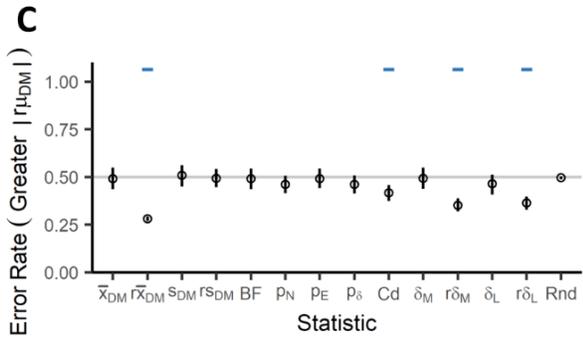
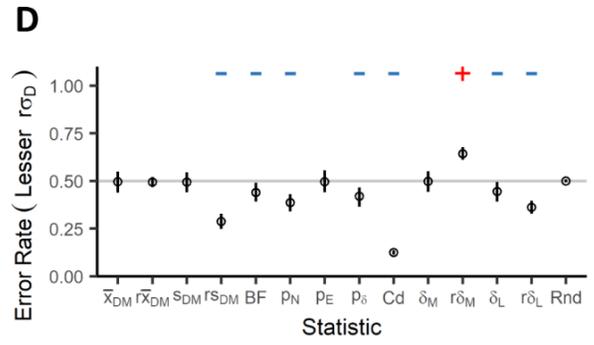
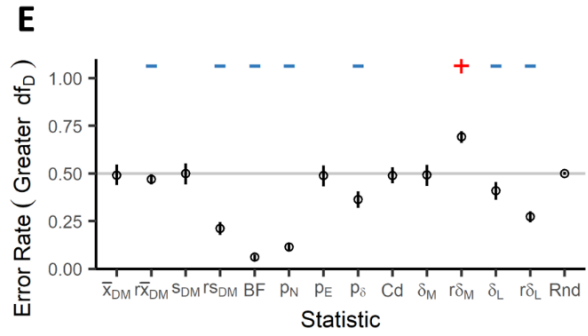
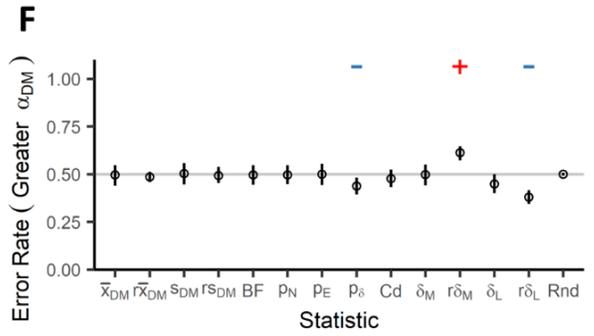

**Fig. S10: $r\delta_L$ is the only statistic that has lower than random comparison error with positive signed statistically significant results across all measures of relative effect strength simultaneously.** (**A**) Fraction of population configurations where experiment 1 has higher effect strength than (HEST) experiment 2 according to each effect strength measure, with $r\mu_{DM}$, $r\sigma_D$, $df_D$, and $\alpha_{DM}$ serving as separate ground truths simultaneously. (**B**) Histogram of the t-ratio for population configurations, indicating that results from experiment 1 (blue) and experiment 2 (pink) are both associated with positive signed statistically significant results ($\bar{t}_{statistic} / |t_{critical}| > 1$). From a single data set, mean comparison error rate of candidate statistics in identifying which experiment has higher relative effect strength via (**C**) higher $r\mu_{DM}$, (**D**) lower $r\sigma_D$, (**E**) higher $df_D$, and (**F**) higher $\alpha_{DM}$ across population configurations. (A) '#' denotes measures that have a nonrandom number of shared designations with each independent measure (listed at top of y-axis) for which experiments are designated with higher effect strength ($p < 0.05$ from Bonferroni corrected two-tailed binomial test for coefficient equal to 0.5 between each independent measure and every effect strength measure). (B) Discrete Kolmogorov-Smirnov test between histograms. (C, D, E, F) Pairwise t-test with Bonferroni correction for all combinations, where blue minus denotes a mean error rate lower than random, red plus denotes higher than random. N=1E3 population configurations generated for each study, n=1E2 samples drawn per configuration, 5 - 30 observations per sample.

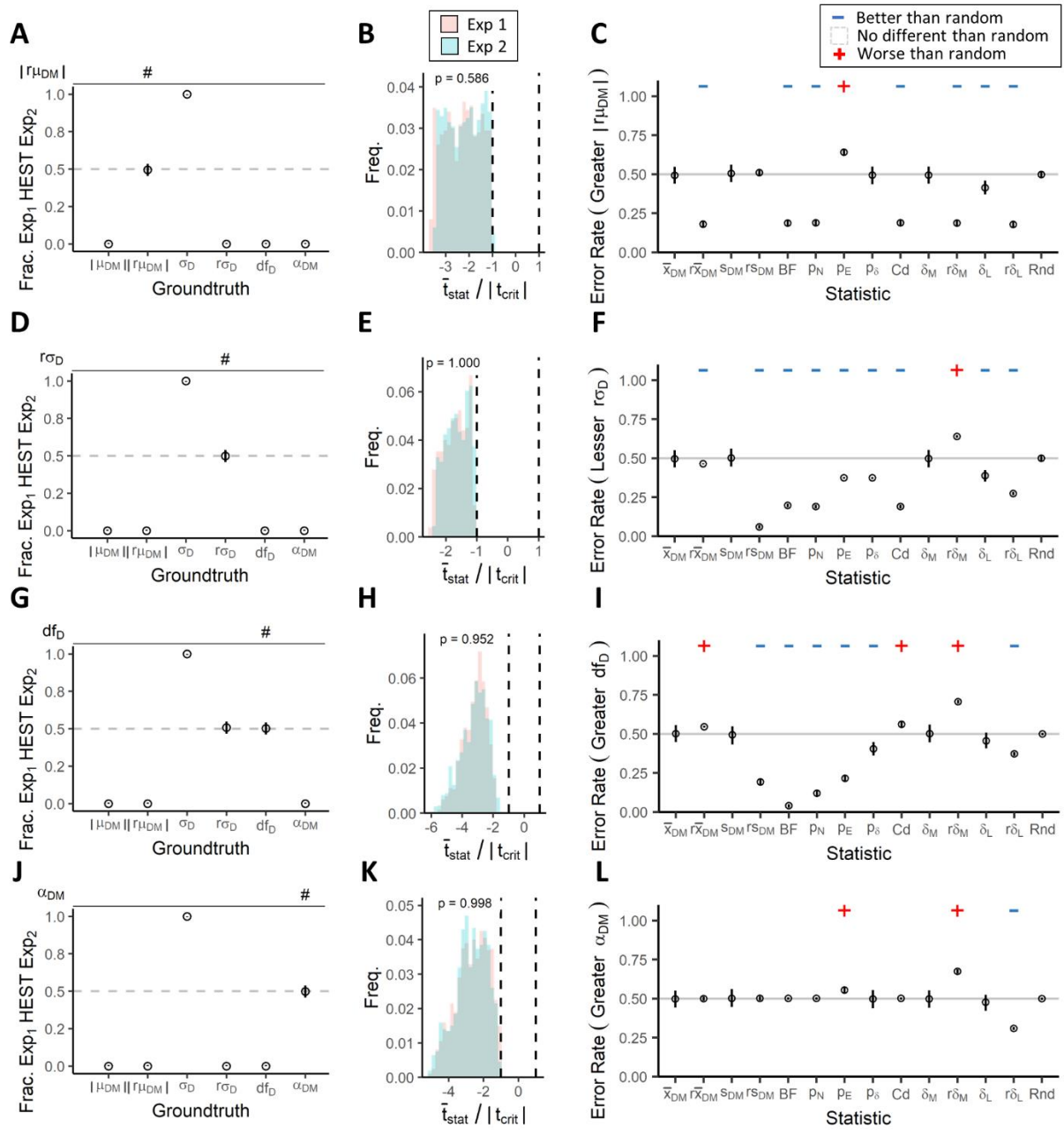

**Fig. S11: $r\delta_L$ is the only statistic that has lower than random comparison error with negative signed statistically significant results for each measure of relative effect strength.** (**A**) Fraction of population configurations where experiment 1 has higher effect strength than (HEST) experiment 2 according to each effect strength measure, with $r\mu_{DM}$ serving as ground truth. (**B**) Histogram of the t-ratio for population configurations, indicating that results from experiment 1 (blue) and experiment 2 (pink) are both associated with negative signed statistically significant results ($\bar{t}_{statistic} / |t_{critical}| < -1$). (**C**) Mean comparison error rate of candidate statistics in identifying which experiment has higher relative effect strength via higher $r\mu_{DM}$ across population configurations (50 observations per sample). (**D**) Fraction of population configurations where experiment 1 has higher effect strength than experiment 2 according to each effect strength measure, with $r\sigma_D$ serving as ground truth. (**E**) Histogram of the t-ratio indicating population configurations are associated with negative signed statistically significant results. (**F**) Mean comparison error rate of candidate statistics in identifying which experiment has higher relative effect strength via lower $r\sigma_D$ across population configurations (50 observations per sample). (**G**) Fraction of population configurations where experiment 1 has higher effect strength than experiment 2 according to each effect strength measure, with $df_D$ serving as ground truth. (**H**) Histogram of the t-ratio indicating population configurations are associated with negative signed statistically significant results. (**I**) Mean comparison error rate of candidate statistics in identifying which experiment has higher relative effect strength via higher $df_D$ across population configurations (6 - 30 observations per sample). (**J**) Fraction of population configurations where experiment 1 has higher effect strength than experiment 2 according to each effect strength measure, with $\alpha_{DM}$ serving as ground truth. (**K**) Histogram of the t-ratio indicating population configurations are associated with negative signed statistically significant results. (**L**) Mean comparison error rate of candidate statistics in identifying which experiment has higher relative effect strength via higher $\alpha_{DM}$ across population configurations (50 observations per sample). (A, D, G, J) '#' denotes measures that have a nonrandom number of shared designations with independent measure (listed at top of y-axis) for which experiments are designated with higher effect strength ($p < 0.05$ from Bonferroni corrected two-tailed binomial test for coefficient equal to 0.5 between independent measure and each effect strength measure). (B, E, H, K) Discrete Kolmogorov-Smirnov test between histograms. (C, F, I, L) Pairwise t-test with Bonferroni correction for all combinations, where blue minus denotes a mean error rate lower than random, red plus denotes higher than random. N=1E3 population configurations generated for each study, n=1E2 samples drawn per configuration.

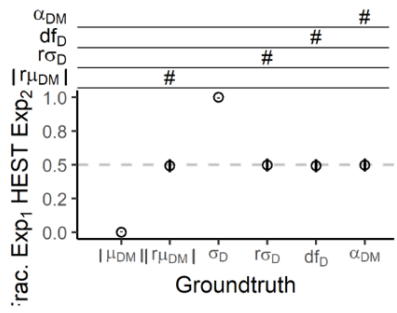
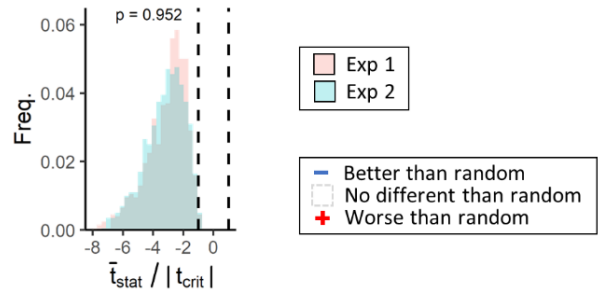
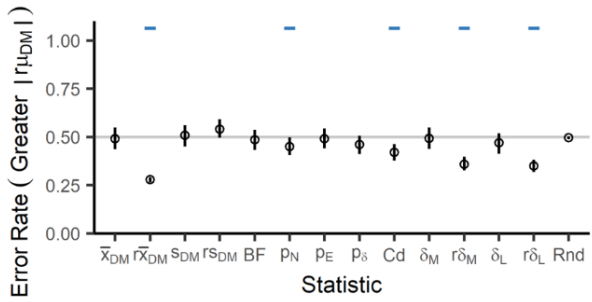
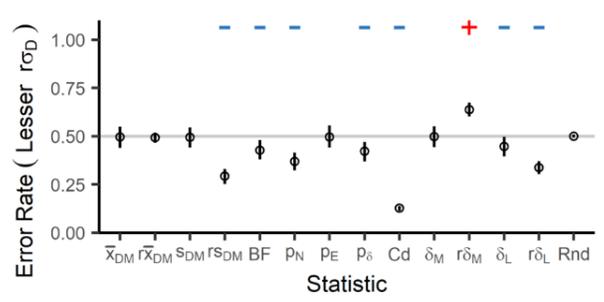
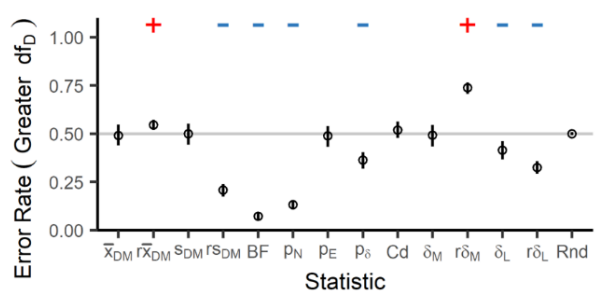
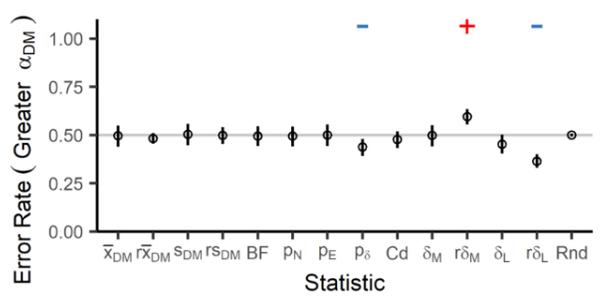

**Fig. S12: $r\delta_L$ is the only statistic that has lower than random comparison error with negative signed statistically significant results across all measures of relative effect strength simultaneously.** (**A**) Fraction of population configurations where experiment 1 has higher effect strength than (HEST) experiment 2 according to each effect strength measure, with $r\mu_{DM}$, $r\sigma_D$, $df_D$, and $\alpha_{DM}$ serving as separate ground truths simultaneously. (**B**) Histogram of the t-ratio for population configurations, indicating that results from experiment 1 (blue) and experiment 2 (pink) are both associated with negative signed statistically significant results ($\bar{t}_{statistic} / |t_{critical}| < -1$). From a single data set, mean comparison error rate of candidate statistics in identifying which experiment has higher relative effect strength via (**C**) higher $r\mu_{DM}$, (**D**) lower $r\sigma_D$, (**E**) higher $df_D$, and (**F**) higher $\alpha_{DM}$ across population configurations. (A) '#' denotes measures that have a nonrandom number of shared designations with each independent measure (listed at top of y-axis) for which experiments are designated with higher effect strength ($p < 0.05$ from Bonferroni corrected two-tailed binomial test for coefficient equal to 0.5 between each independent measure and every effect strength measure). (B) Discrete Kolmogorov-Smirnov test between histograms. (C, D, E, F) Pairwise t-test with Bonferroni correction for all combinations, where blue minus denotes a mean error rate lower than random, red plus denotes higher than random. N=1E3 population configurations generated for each study, n=1E2 samples drawn per configuration, 6 - 50 observations per sample.

**Table S1:** Candidate Summary Statistics to Evaluate Comparison Error for Effect Strength

| Statistic | Equation | Decision Rule |
|---|---|---|
| $\bar{x}_{DM}$ | $\bar{y} - \bar{x}$ | $|\bar{x}_{DM,1}| > |\bar{x}_{DM,2}|$ |
| $r\bar{x}_{DM}$ | $\dfrac{\bar{x}_{DM}}{\bar{x}}$ | $|\bar{x}_{DM,1}| > |\bar{x}_{DM,2}|$ |
| $s_{DM}$ | $\sqrt{\dfrac{s_X^2}{m} + \dfrac{s_Y^2}{n}}$ | $|s_{DM,1}| < |s_{DM,2}|$ |
| $rs_{DM}$ | $\dfrac{s_{DM}}{\bar{x}}$ | $|rs_{DM,1}| < |rs_{DM,2}|$ |
| BF (2) | $\dfrac{Pr(D|M_1)}{Pr(D|M_2)}$ | $BF_1 > BF_2$ |
| $p_N$ (3) | $P\left(Z \geq \dfrac{\bar{y} - \bar{x}}{\sqrt{\sigma_X^2/m + \sigma_Y^2/n}}\right)$ | $p_{NHST,1} < p_{NHST,2}$ |
| $p_E$ (4) | $Max\left\{\left(\Delta \leq \dfrac{\bar{y} - \bar{x} - \Delta_L}{\sqrt{\sigma_X^2/m + \sigma_Y^2/n}}\right), \left(\Delta \geq \dfrac{\bar{y} - \bar{x} + \Delta_U}{\sqrt{\sigma_X^2/m + \sigma_Y^2/n}}\right)\right\}$ | $p_{TOST,1} > p_{TOST,2}$ |
| $p_\delta$ (5) | $p_\delta = \dfrac{|I \cap H_0|}{|I|} \times max\left\{\dfrac{|I|}{2|H_0|}, 1\right\}$ | $p_{\delta,1} < p_{\delta,2}$ |
| CD (6) | $\dfrac{\bar{y} - \bar{x}}{\sqrt{\dfrac{(m-1)s_X^2 + (n-1)s_Y^2}{m+n-2}}}$ | $|CD_1| > |CD_2|$ |
| $\delta_M$ | $N^{-1}\sum_{i=1}^{K}\mathbb{I}(\mu_Y^i - \mu_X^i \leq c) - N^{-1}\sum_{i=1}^{K}\mathbb{I}(\mu_Y^i - \mu_X^i \leq -c) = 1 - \alpha_{DM}$ | $\delta_{M,1} > \delta_{M,2}$ |
| $r\delta_M$ | $N^{-1}\sum_{i=1}^{K}\mathbb{I}\left(\dfrac{\mu_Y^i - \mu_X^i}{\mu_X^i} \leq c\right) - N^{-1}\sum_{i=1}^{K}\mathbb{I}\left(\dfrac{\mu_Y^i - \mu_X^i}{\mu_X^i} \leq -c\right) = 1 - \alpha_{DM}$ | $r\delta_{M,1} > r\delta_{M,2}$ |
| $\delta_L$ (1) | $sign(\bar{x}_{DM})\left(sign(b_{lo}) == sign(b_{hi})\right) min(|b_{lo}|, |b_{hi}|)$ | $\delta_{L,1} < \delta_{L,2}$ |
| $r\delta_L$ (1) | $sign(r\bar{x}_{DM})\left(sign(c_{lo}) == sign(c_{hi})\right) min(|c_{lo}|, |c_{hi}|)$ | $r\delta_{L,1} > r\delta_{L,2}$ |
| Rnd | | $Rnd_1 < Rnd_2$ |

*Note: decision rule is logical expression that predicts experiment 1 has higher effect strength than experiment 2 when true.*

*Abbreviations: $\bar{x}$, sample mean of control group; $\bar{y}$, sample mean of experiment group; $s_X$, sample standard deviation of control group; $s_Y$, sample standard deviation of experiment group; $\bar{x}_{DM}$, difference in sample means; $r\bar{x}_{DM}$, relative difference in sample means; $s_{DM}$, standard deviation of the difference in sample means; $rs_{DM}$, relative standard deviation of the difference in sample means; BF, Bayes Factor; $p_{NHST}$, p-values from null hypothesis significance test; $p_{TOST}$, p-value from two one sided t-tests; $p_\delta$, second generation p-value; CD, cohen's d; $\delta_M$, most difference in means;*

*rδ<sub>M</sub>, relative most mean difference in sample means; δ<sub>L</sub>, least difference in means; rδ<sub>L</sub>, relative least difference in means; Rnd, random.*

**Table S2:** Loss functions for Each Measure of Effect Strength

| Measure | Loss Functions: $Loss(x, y, x', y', \theta, \theta') :=$ | Eq. |
|---|---|---|
| $|\mu_{DM}|$: | $1 - \mathbb{I}(|\mu_{DM}| > |\mu'_{DM}|$ and $|\delta(x, y, \alpha_{DM})| < |\delta(x', y', \alpha'_{DM})|)$ | (S8) |
| $\sigma_D$: | $1 - \mathbb{I}(\sigma_D < \sigma'_D$ and $|\delta(x, y, \alpha_{DM})| < |\delta(x', y', \alpha'_{DM})|)$ | (S9) |
| $df_D$: | $1 - \mathbb{I}(df_D > df'_D$ and $|\delta(x, y, \alpha_{DM})| < |\delta(x', y', \alpha'_{DM})|)$ | (S10) |
| $\alpha_{DM}$: | $1 - \mathbb{I}(\alpha_{DM} > \alpha'_{DM}$ and $|\delta(x, y, \alpha_{DM})| < |\delta(x', y', \alpha'_{DM})|)$ | (S11) |
| $|r\mu_{DM}|$: | $1 - \mathbb{I}(|r\mu_{DM}| > |r\mu'_{DM}|$ and $|\delta(x, y, \alpha_{DM})| < |\delta(x', y', \alpha'_{DM})|)$ | (S12) |
| $r\sigma_D$: | $1 - \mathbb{I}(r\sigma_{DM} < r\sigma'_{DM}$ and $|\delta(x, y, \alpha_{DM})| < |\delta(x', y', \alpha'_{DM})|)$ | (S13) |

*Note: decision rule for candidate prediction (δ) may by a "greater than" or "less than" operation depending on candidate statistic. The loss functions specifies when the prediction disagrees with the ground truth designation. In this case, the ground truth designations are for higher effect strength for experiment 1 vs experiment 2 ('), and the candidate predictions test for lower effect strength for experiment 1.*

**Table S3:** Positive Results for Total Plasma Cholesterol

| | r$\bar{x}_{DM}$ | Group A | $\bar{x}$ | $s_X$ | m | Group B | $\bar{y}$ | $s_Y$ | n | Units | $\alpha_{DM}$ | Sp | PMID, Loc |
|---|---|---|---|---|---|---|---|---|---|---|---|---|---|
| 1 | -30% | Vehicle | 1335 | 269 | 8 | Atorvastin | 934 | 232 | 8 | mg/dL | 0.05/6 | rb | 24188322, F1 |
| 2 | -33% | Placebo | 202 | 12.6 | 5 | 150 mg Atorvastatin | 135.2 | 28.7 | 5 | mg/dL | 0.05/6 | hu | 22716983, ST9 |
| 3 | -21% | ApoE$^{-/-}$ PAI-1$^{WT}$ | 2503 | 266 | 11 | ApoE$^{-/-}$ PAI-1$^{-/-}$ | 1984 | 252 | 13 | mg/dl | 0.05 | ms | 10712412, T1 |
| 4 | -45% | Saline + WTD | 463 | 103 | 12 | PCSK9-mAb1 + WTD | 254 | 108 | 10 | mg/dl | 0.05/21 | ms | 31366894, F1A |
| 5 | -29% | Ldlr$^{-/-}$Ad-Gal-Reln$^{FL/FL}$ | 2087 | 531 | 15 | Ldlr$^{-/-}$ Ad-Cre-Reln$^{FL/FL}$ | 1487 | 364 | 16 | mg/dl | 0.05 | ms | 26980442, SF2B |
| 6 | -36% | Luciferase siSRNA | 238 | 15.7 | 5 | Angptl3 siRNA | 153 | 22.4 | 5 | mg/dL | 0.05/6 | ms | 32808882, F1D |
| 7 | -31% | WTD -IF | 4.78 | 1.21 | 20 | WTD +IF | 3.3 | 0.67 | 20 | mmol/L | 0.05/4 | ms | 9614153, F1A |
| 8 | -31% | HFD -IF | 4.78 | 1.21 | 20 | HFD +IF | 3.3 | 0.67 | 20 | mmol/L | 0.05/4 | ms | 9614153, F1A |
| 9 | -45% | WTD +Saline | 463 | 103 | 12 | WTD +PCSK9-mAb1 | 254 | 108 | 10 | mg/dl | 0.05 | ms | 31366894, F1A |
| 10 | -52% | WT | 95.7 | 9.4 | 4 | Pcsk9$^{-/-}$ | 46.3 | 1.9 | 4 | mg/dL | 0.05 | ms | 15805190, T1 |
| 11 | -58% | ApoE$^{-/-}$ | 300 | 97 | 13 | ApoE$^{-/-}$ +Palm-E | 126 | 41 | 14 | mg/dl | 0.05 | ms | 11015467, T1 |
| 12 | -56% | pCMV5 | 642 | 63 | 9 | PCMV-E3 | 283 | 69 | 10 | mg/dl | 0.05 | ms | 11110410, F3 |
| 13 | -69% | HFD -Ezetimibe | 268 | 53.8 | 6 | HFD +Ezetimibe | 82.3 | 30.7 | 10 | mg/dL | 0.05/5 | mk | 11245855, F4 |
| 14 | -82% | HFD -Simvastin | 241 | 10 | 10 | HFD +Simvastin | 44 | 15 | 5 | mg/dL | 0.05/6 | hm | 9162756, T2 |

**Table S4:** Positive Results for Total Plasma Cholesterol

| | r$\bar{x}_{DM}$ | Group A | $\bar{x}$ | $s_X$ | m | Group B | $\bar{y}$ | $s_Y$ | n | Units | $\alpha_{DM}$ | Sp | PMID, Loc |
|---|---|---|---|---|---|---|---|---|---|---|---|---|---|
| 1 | -51% | Vehicle | 304788 | 1E+05 | 4 | SC-69000 | 149779 | 34576 | 7 | µm$^2$ | 0.05/2 | pg | 10571535, T2 |
| 2 | -52% | ApoE$^{-/-}$ | 46.2 | 10.6 | 3 | ApoE$^{WT}$ | 22.1 | 8 | 3 | % | 0.05 | pg | 30305304, F5A |
| 3 | -28% | ApoE$^{-/-}$ | 225000 | 72732 | 10 | ApoE$^{-/-}$ + PAO | 161000 | 47434 | 10 | µm$^2$ | 0.05 | ms | 27683551, F5C |
| 4 | -55% | ApoE$^{-/-}$Cyp17a1$^{WT}$ WTD♀ | 2.68 | 1.151 | 6 | ApoE$^{-/-}$Cyp17a1$^{-/-}$ WTD♀ | 1.2 | 0.805 | 5 | % | 0.05 | ms | 32472014, F3B |
| 5 | -24% | ApoE$^{-/-}$ | 0.54 | 0.12 | 16 | ApoE$^{-/-}$P2Y$_1$$^{-/-}$ | 0.41 | 0.116 | 15 | mm$^2$ | 0.05 | ms | 18663083, F2B |
| 6 | -47% | ApoE$^{-/-}$ | 16.1 | 7.7 | 10 | ApoE$^{-/-}$EC-TFEB | 8.58 | 3.3 | 12 | % | 0.05 | ms | 28143903, F7F |
| 7 | -69% | Progression Cotrol | 0.713 | 0.297 | 8 | Atorvastin | 0.221 | 0.147 | 8 | mm$^2$ | 0.05/7 | rb | 7840808, T4 |
| 8 | -48% | WT WTD -IF | 3.1 | 1 | 16 | WT WTD +IF | 1.6 | 1.0 | 20 | mm$^2$ | 0.05 | ms | 9614153, F2B |
| 9 | -73% | LDLr$^{-/-}$ -IF | 7.39 | 6.7 | 13 | LDLr$^{-/-}$ +IF | 2.01 | 3.4 | 15 | % | 0.05 | ms | 9614153, F2B |
| 10 | -91% | Vehicle | 0.173 | 0.150 | 8 | Probucol | 0.015 | 0.025 | 8 | mm$^2$ | 0.05/6 | rb | 24188322, F1 |
| 11 | -66% | Ldlr$^{-/-}$ Fxr$^{WT}$ ♂ | 42.4 | 19.8 | 13 | Ldlr$^{-/-}$ Fxr$^{-/-}$ ♂ | 14.47 | 10.2 | 13 | % | 0.05/2 | ms | 16825595, F3B |
| 12 | -82% | Distal Aorta +Stenosis | 39 | 46 | 15 | Distal Aorta +Stenosis | 7 | 11 | 13 | % | 0.05 | mk | 3795393, T3 |
| 13 | -73% | ApoE$^{-/-}$ BMT ApoE$^{-/-}$ | 472 | 118 | 6 | ApoE$^{-/-}$ BMT ApoE$^{+/+}$ | 126 | 43 | 6 | | 0.05 | ms | 7863332, T1 |
| 14 | -91% | Saline | 0.68 | 0.17 | 9 | LDE-etoposide | 0.06 | 0.06 | 9 | % | 0.05 | ms | 22072867, T3 |

**Supplement References**